\documentclass[epj]{svjour}
\usepackage{amsmath,amssymb}
\usepackage{times}
\usepackage{txfonts}
\usepackage{graphicx,color}
\usepackage{dcolumn}
\usepackage{bm}

\newcommand{\natu}{$^\mathrm{nat}$}
\newcommand{\del}{\mathrm{d}}
\usepackage{lscape}

\DeclareSymbolFont{EulerExtension}{U}{euex}{m}{n}
\DeclareMathSymbol{\euintop}{\mathop} {EulerExtension}{"52}
\DeclareMathSymbol{\euointop}{\mathop} {EulerExtension}{"48}

\everymath{\displaystyle}

\begin{document}
\title{
Isomer production studied with simultaneous decay curve analysis for alpha-particle induced reactions on natural platinum up to 29~MeV
}

\author{
Naohiko Otuka\inst{1.2}\thanks{n.otsuka@iaea.org}
\and
S\'{a}ndor Tak\'{a}cs\inst{3}\thanks{stakacs@atomki.hu}
\and
Masayuki Aikawa\inst{2,4,5,6}\thanks{aikawa@sci.hokudai.ac.jp}
\and
Shuichiro Ebata\inst{7}
\and
Hiromitsu Haba\inst{2}
}
\institute{
Nuclear Data Section,
Division of Physical and Chemical Sciences,
Department of Nuclear Sciences and Applications,
International Atomic Energy Agency,
A-1400 Wien, Austria
\and
Nishina Center for Accelerator-Based Science,
RIKEN,
Wako 351-0198,
Japan
\and
HUN-REN Institute for Nuclear Research (ATOMKI),
H-4026 Debrecen,
Hungary
\and
Faculty of Science,
Hokkaido University,
Sapporo 060-0810,
Japan
\and
Graduate School of Biomedical Science and Engineering,
Hokkaido University,
Sapporo 060-8638,
Japan
\and
Global Center for Biomedical Science and Engineering,
Faculty of Medicine,
Hokkaido University,
Sapporo 060-8648,
Japan
\and
Graduate School of Science and Engineering,
Saitama University,
Saitama 338-8570,
Japan
}
\date{Received: date / Revised version: date}

\abstract{
The isomeric ratios of $^{198}$Au, $^{197}$Hg and $^{195}$Hg produced by $\alpha$-particle induced reactions on natural platinum were investigated experimentally up to 29~MeV by using the standard stacked foil activation technique and $\gamma$-ray spectrometry.
The isomeric ratios of $^{197}$Hg and $^{195}$Hg determined by the conventional activation cross section formula showed strong cooling time dependence.
The time dependence was resolved by adjusting the isomeric transition branching ratios for the two isotopes within a simultaneous decay curve analysis framework.
Our analysis suggests 94.5$\pm$0.7\% and 48.9$\pm$1.8\% as the isomeric transition branching ratios of $^{197m}$Hg (24~h) and $^{195m}$Hg (42~h),
respectively.
The isomeric ratios and independent production cross sections of $^{198}$Au, $^{197}$Hg, $^{195}$Hg and some other Hg, Au and Pt isotopes were also measured down to 6~MeV with these corrected isomeric transition branching ratios,
and compared with predictions of statistical and pre-equilibrium models by TALYS-2.0 to discuss spin cutoff parameter dependence.
We found the measured isomeric ratios are better predicted if we reduce the spin cutoff parameter to half or less from that estimated with the rigid body moment of inertia.
}

\maketitle
\onecolumn
\section{Introduction}
Knowledge of the isomeric ratio in a nuclear process is important for both theoretical and experimental nuclear physics (e.g., spin dependence of level density) and nuclear applications (e.g., radioisotope production, nuclear waste management)~\cite{Rodrigo2023}.
In addition to the general purpose experimental nuclear reaction data library EXFOR~\cite{Otuka2014},
compilation specialized to the experimental isomeric ratios has been made for fission products~\cite{Sears2021} and reaction products~\cite{Rodrigo2023}.

An isomeric ratio is typically measured by the ground state and metastable state activities of a sample after irradiation if both states have radiations and half-lives suitable for detection.
Quantification of the metastable state production cross section $\sigma_m$ is usually straightforward as long as the state is shielded against decay of the neighboring nuclides and the metastable state has at least one $\gamma$ line not related with the decay of the ground state.
Quantification of ground state production could be more challenging because it may require subtraction of the portion originating from the isomeric transition from the metastable state production.

When a product nuclide $i$ may be formed directly by a reaction and indirectly by decay from a precursor product nuclide \textit{1}, \textit{2}, etc. via \textit{1} $\to$ \textit{2} $\to$ $\cdots$ $\to$ $i$,
the cross section for production of the nuclide $i$ is related with the production cross sections of the precursor nuclides $j$ ($j=1,i-1$) by
\begin{equation}
\sigma_i=\sigma_i^\mathrm{cum} 
        -\frac{\lambda_i}{f_i}
        \sum_{j=1}^{i-1}\sigma_jp_{ji}
        \sum_{k=j}^i \Lambda_k^{j,i}\frac{f_k}{\lambda_k},
\label{eqn:cumgen}
\end{equation}
where
\begin{equation}
\Lambda_k^{j,i}=
\left\{
\begin{array}{ll}
1
& \mathrm{if\quad} i=j
\\
\left.
\prod_{l=j}^{i-1}\lambda_l
\right/
\prod_{\substack{l=j\\(l\ne k)}}^i(\lambda_l-\lambda_k)
&\mathrm{otherwise}
\end{array}
\right.
,
\label{eqn:cfunction}
\end{equation}
$p_{ji}$ is the branching ratio for decay of the nuclide $j$ to $i$,
$\lambda_k$ is the decay constant of the nuclide $k$,
$f_k = [1 - \exp(-\lambda_k t_b)]
            \exp(-\lambda_k t_c)
       [1 - \exp(-\lambda_k t_m)]/\lambda_k$
with the bombardment time $t_b$, cooling time $t_c$ and measurement time $t_m$.\footnote{
See Appendix~\ref{sec:app7} for its proof.
}
The symbol $\sigma_i^\mathrm{cum}$ denotes the production cross section including precursor decay contribution,\footnote{
The cross section $\sigma_i^\mathrm{cum}$ is not explicitly referred as ``cumulative cross section" through this manuscript since it is not uniquely defined in the community~\cite{David2015}. 
}
and it is expressed by the usual activation cross section formula $\sigma_i^\mathrm{cum} = C_i / (f_i \phi n)$, where $C_i$ is the number of decays of the nuclide $i$ during the measurement time, $\phi$ is the beam flux and $n$ is the sample atom areal number density.

If the nuclide $i$ is a ground state ($g$) and only its metastable state ($m$) may decay into the ground state with the branching ratio $p$,
Eq.~(\ref{eqn:cumgen}) is simplified to
\begin{equation}
\sigma_g=\sigma_g^\mathrm{cum}
        -p\left[
         \frac{f_m}{f_g}\frac{\lambda_g}{\lambda_g-\lambda_m}
        -\frac{\lambda_m}{\lambda_g-\lambda_m}
         \right]\sigma_m,
\label{eqn:cumgm1}
\end{equation}
as demonstrated in~\cite{Vanska1981}.
Equation~(\ref{eqn:cumgm1}) shows $\sigma_g^\mathrm{cum}\ne\sigma_g+p\sigma_m$ in general.
It is often not trivial to know which one is reported by the experimentalist,
and both $\sigma_g^\mathrm{cum}$ and $\sigma_g+p\sigma_m$ are compiled in the EXFOR library just as the production cross section of the nuclide when $p$=1.
For example,
the cross sections for production of $^{198}$Au determined by measurement of the $^{198}$Au ground state activity for a platinum foil irradiated by an $\alpha$-particle beam is noted by \texttt{(78-PT-0(A,X)79-AU-198,,SIG)} rather than \texttt{(78-PT-0(A,X)79-AU-198-G,CUM,SIG)} or \texttt{(78-PT-0(A,X)79-AU-198-G+M,,SIG)} in the EXFOR library.

When the half-life of the metastable state is shorter than the half-life of the ground state ($\lambda_m > \lambda_g$) and the cooling time is enough longer than the half-life of the metastable state ($t_c \gg 1/\lambda_m$),
Eq.~(\ref{eqn:cumgm1}) is simplified to
\begin{equation}
\sigma_g=\sigma_g^\mathrm{cum}-p\frac{\lambda_m}{\lambda_m-\lambda_g}\sigma_m,
\label{eqn:cumgm2}
\end{equation}
which is seen in various articles reporting ground state production cross sections.
Equation~(\ref{eqn:cumgm2}) can be further simplified to
\begin{equation}
\sigma_g=\sigma_g^\mathrm{cum}-p\sigma_m
\label{eqn:cumgm3}
\end{equation}
if $\lambda_m\gg\lambda_g$.
According to systematic conversion of isomer production cross sections in the EXFOR library to their isomeric ratios~\cite{Rodrigo2023},
we noticed that the experiments aimed at determining $\sigma_g$ could be less attentive to the conditions under which Eqs.~(\ref{eqn:cumgm2}) and (\ref{eqn:cumgm3}) are valid though some publications reporting production cross sections discuss subtraction of precursor contribution in details with proper equations (e.g.,~\cite{Titarenko2002,Lebeda2010}).

By definition,
the reaction cross section and isomeric ratio are cooling-time independent quantities.
However, one may obtain cooling-time dependent $\sigma_g$ if Eq.~(\ref{eqn:cumgm2}) is applied to a $\gamma$ spectrum obtained without sufficiently long cooling time and $f_m/f_g$ in Eq.~(\ref{eqn:cumgm1}) does not reach its saturation value.
The cooling time dependence also may be caused by interference due to the presence of the same $\gamma$ line following $\beta$ decay or electron capture of the metastable state and ground state,
or due to the presence of a similar $\gamma$ line attributed to a co-produced nuclide not separable due to finite resolution of the detector.
Use of an incorrect decay branching ratio of the metastable state can be another reason to see the cooling time dependence.

An irradiation of natural platinum by an $\alpha$-particle beam may produce several radionuclide pairs of the ground and metastable states such as $^{198g}$Au (2.7~d) and $^{198m}$Au (2.3~d), $^{197g}$Hg (64~h) and $^{197m}$Hg (24~h), and $^{195g}$Hg (11~h) and $^{195m}$Hg (42~h).
The two states of $^{198}$Au have very similar half-lives and could be a good example showing that use of Eq.~(\ref{eqn:cumgm3}) is inappropriate ($p \lambda_m/(\lambda_m-\lambda_g)\sim 6.38$).
A variety of the energy dependence seen in the experimental isomeric ratios of the \natu Pt($d$,x)$^{198}$Au reaction in the literature ~\cite{Ditroi2006,Khandaker2015,Tarkanyi2019} could be discussed in this context.
The difference of the ground state and metastable state half-lives is larger for $^{197}$Hg ($p \lambda_m/(\lambda_m-\lambda_g)\sim 1.45$),
but the measured isomeric ratio of the $^{197}$Au($d,2n$)$^{197}$Hg reaction in the literature still exhibit large discrepancy~\cite{Rodrigo2023}.
Measurements of the isomeric ratio of the \natu Pt($\alpha$,x)$^{197}$Hg reaction have been also adopted for theoretical modelling of the isomeric ratio by introducing the spin cutoff parameter~\cite{Vandenbosch1960,Sudar2006}.
Measurements of the isomeric ratio for the $^{195}$Hg pair should take into account the fact that the half-life of the metastable state is longer than the half-life of the ground state,
namely Eq.~(\ref{eqn:cumgm2}) is not usable at all.
These three pairs of the isomers are shielded from decay of the neighboring nuclides on the same mass chain under irradiation of a platinum foil by an $\alpha$-particle beam,
and this production route is ideal to study the isomeric ratios of these nuclides.
Among the three nuclides mentioned above, 
$^{198g}$Au and $^{197m,g}$Hg also have been considered as therapeutic radionuclides~\cite{Tarkanyi2024},
and their production cross sections are important for medical application.

The purpose of the present work is to discuss the origin of the ``unphysical" cooling time dependence seen in our isomeric ratio measurement and to discuss
the importance of adopting proper data reduction procedures and \textit{obtaining} decay data with $\gamma$-ray measurements repeated throughout a sufficiently long cooling period.

\section{Experimental}
Pure metallic foils of natural platinum (2.53$\times$2.56~cm, 0.0804~g, 12.433~mg/cm$^2$), natural titanium (5.02$\times$10.03~cm, 0.1180~g, 2.344~mg/cm$^2$) and aluminium (4.93$\times$4.97~cm, 0.03672~g, 1.500~mg/cm$^2$) were cut into 8$\times$8~mm to form a target stack.
The Ti foils were interleaved to examine the beam flux and energy by using the \natu Ti($\alpha$,x)$^{51}$Cr reaction,
while the Al foils were interleaved to catch the nuclei produced in the adjacent Pt foil and recoiled.
The molar masses recommended by IUPAC~\cite{Meija2016} and adopted by us are 195.08~g/mol for Pt and 47.867~g/mol for Ti.
Nine Pt, nine Ti foils and eighteen Al foils were arranged in a stack of nine sets of Pt-Al-Ti-Al,
which were followed by four additional Ti foils.
They were stacked in a target holder,
which also served as a Faraday cup.
The incoming and outgoing beam energies at each foil were calculated by using the SRIM code~\cite{Ziegler2010}.

The target stack was irradiated for 60~min by an $\alpha$-particle beam extracted from the RIKEN AVF cyclotron and collimated to 3~mm in diameter.
The initial beam energy was determined as 29.0~MeV by the time-of-flight method~\cite{Watanabe2014}.
We monitored the time profile of the beam current by recording the charge integrated in the Faraday cup periodically,
and confirmed that the beam flux was stable during irradiation with the average beam intensity of 6.20$\times 10^{11}$ particles/sec ($\sim$199~nA).

Measurements of the activities of the irradiated foils were started about 1~h after the end of bombardment using a high-purity germanium detector (ORTEC GEM30P4-70) connected to a software tool (SEIKO EG\&G Gamma Studio).
The detection efficiency was determined at various $\gamma$ energies by using a calibrated multiple point-like source consisting of $^{57,60}$Co, $^{85}$Sr, $^{88}$Y, $^{109}$Cd, $^{113}$Sn $^{137}$Cs, $^{139}$Ce, $^{203}$Hg and $^{241}$Am at each distance between the detector surface and sample position.
We assumed that the detection efficiency $\epsilon$ is related with the $\gamma$ energy $E$ by 
\begin{equation}
\ln \epsilon=\sum_{i=0}^5 a_i\left(\ln E\right)^i,
\end{equation}
and the coefficients $a_i$ ($i=0,5$) were adjusted to reproduce the measured detection efficiencies at the calibration energies.
Correction for absorption of $\gamma$-rays in the foils was done only for the 98 keV $\gamma$ line peak areas with the mass attenuation coefficient $\mu/\rho$=5.133~cm$^2$/g~\cite{Hubbell2004},
which introduces attenuation by 3.124\%.

The activity of the first pair of the Pt and Al foils at the highest energy position in the stack was measured together more than 50 times.
The measurements were repeated for about 10 days after the end of bombardment to study the cooling time dependence of the measured isomeric ratios.
The activities of the rest of the Pt and Al foil pairs and the Ti monitor foils were measured four times or less.
Additionally,
the activities of the first and second pairs of Pt and Al foils were measured about 10~months after the bombardment to study the activity of the long-lived $^{195}$Au (186~d).
In the present work,
isomeric ratios and production cross sections were derived from the measured $\gamma$-ray counts, detection efficiencies, irradiation and target parameters as well as the decay data taken from the Evaluated Nuclear Structure and Data File (ENSDF) library~\cite{A200,A199,A198,A197,A196,A195,A194,A192,A51} via the IAEA LiveChart of Nuclides~\cite{Verpelli2011} as summarized in Table~\ref{tab:decaydata}.
%
%
Considering the time scale of our measurement,
we do not treat an excitation level with its half-life shorter than 1~min as a metastable state.
\begin{table}[hbtp]
\caption{
Half-life $T_{1/2}$, decay branching ratio $p$, $\gamma$ energy $E_\gamma$ and $\gamma$ emission probability $I_\gamma$ adopted in the present work taken from the ENSDF library~\cite{A200,A199,A198,A197,A196,A195,A194,A192,A51}.
``Route" gives the production route via a reaction having the lowest threshold energy as well as decay.
All $\gamma$ lines listed here were used in determination of the cross sections and isomeric ratios by the simultaneous decay curve analysis.
The $\gamma$ energy in the ENSDF library is rounded to one digit after the decimal point.
The threshold energy $E_\mathrm{thr}$ was calculated by Tcalc~\cite{Shimizu2024} with the atomic masses in AME2020~\cite{Huang2021}.
The parenthesized number is the uncertainty corresponding to the last digits of the estimate,
e.g., 0.52(10) means 0.52$\pm$0.10. 
}
\label{tab:decaydata}
\begin{center}
\begin{tabular}{lllD{.}{.}{1}lrll}
\hline
\hline
Nuclide    &$T_{1/2}$     &$p$ (\%)              &\multicolumn{1}{l}{$E_\gamma$ (keV)}&$I_\gamma$ (\%) &$E_\mathrm{thr}$ (MeV)&Route                        &Ref.  \\
\hline                                                                                                                                 
$^{200m}$Au&18.7 (5) h    &IT 16 (1)             &133.2           &0.52 (10)       &14.7                  &$^{198}$Pt($\alpha,d$)       &\cite{A200}\\
           &              &                      &332.8           &2.2 (5)         &                      &                             &      \\
\cline{3-5}                                                                                                                            
           &              &$\beta^-$ 84 (1)      &255.9           &72.4 (24)       &                      &                             &      \\
           &              &                      &367.9           &79.3 (9)        &                      &                             &      \\
           &              &                      &579.3           &82.4 (10)       &                      &                             &      \\
\cline{1-7}                                                                                                                            
$^{200g}$Au&48.4 (3) min  &$\beta^-$ 100         &367.9           &19 (3)          &13.6                  &$^{198}$Pt($\alpha,d$)       &      \\
           &              &                      &                &                &                      &$^{200m}$Au (IT)             &      \\
\hline                                                                                                                                 
$^{199m}$Hg&42.67 (9) min &IT 100                &158.3           &52.3 (10)       &1.4                   &$^{195}$Pt($\alpha,\gamma$)  &\cite{A199}\\
           &              &                      &374.1           &13.8 (11)       &                      &                             &      \\
\cline{1-7}                                                                                                                            
$^{199}$Au &3.139 (7) d   &$\beta^-$ 100         &158.4           &40.0 (7)        &8.6                   &$^{196}$Pt($\alpha,p$)       &      \\
           &              &                      &208.2           &8.72 (18)       &                      &                             &      \\
\hline                                                                                                                                 
$^{198m}$Au&2.272 (16) d  &IT 100                &180.3           &49 (5)          &9.1                   &$^{195}$Pt($\alpha,p$)       &\cite{A198}\\
           &              &                      &204.1           &39 (5)          &                      &                             &      \\
           &              &                      &214.9           &77.0 (10)       &                      &                             &      \\
           &              &                      &333.8           &18 (4)          &                      &                             &      \\
\cline{1-7}                                                                                                                            
$^{198g}$Au&2.6941 (2) d  &$\beta^-$ 100         &411.8           &95.62 (6)       &8.2                   &$^{195}$Pt($\alpha,p$)       &      \\
           &              &                      &                &                &                      &$^{198m}$Au (IT)             &      \\
\hline                                                                                                                                 
$^{197m}$Hg&23.8 (1) h    &IT 91.4 (7)           &134.0           &33.5 (3)        &10.4                  &$^{194}$Pt($\alpha,n$)       &\cite{A197}\\
\cline{3-5}                                                                                                                            
           &              &EC 8.6 (7)            &279.0           &6.1 (5)         &                      &                             &      \\
\cline{1-7}                                                                                                                            
$^{197g}$Hg&64.14 (5) h   &EC 100                &191.4           &0.632 (22)      &10.1                  &$^{194}$Pt($\alpha,n$)       &      \\
           &              &                      &                &                &                      &$^{197m}$Hg (IT)             &      \\
\cline{1-7}                                                                                                                            
$^{197m}$Pt&95.41 (18) min&IT 96.7 (4)           &346.5           &11.1 (3)        &8.1                   &$^{198}$Pt($\alpha,n+\alpha$)&      \\
\cline{1-7}                                                                                                                            
$^{197g}$Pt&19.8915 (19) h&$\beta^-$ 100         &191.4           &3.7 (4)         &7.7                   &$^{198}$Pt($\alpha,n+\alpha$)&      \\
           &              &                      &                &                &                      &$^{197m}$Pt (IT)             &      \\
\hline                                                                                                                                 
$^{196m}$Au&9.6 (1) h     &IT 100                &147.8           &43.5 (15)       &15.1                  &$^{195}$Pt($\alpha,t$)       &\cite{A196}\\
           &              &                      &188.3           &30.0 (15)       &                      &                             &      \\
\cline{1-7}                                                                                                                           
$^{196g}$Au&6.1669 (6) d  &EC/$\beta^+$ 93.0 (3) &333.0           &22.9 (9)        &14.5                  &$^{195}$Pt($\alpha,t$)       &      \\
           &              &                      &355.7           &87 (3)          &                      &$^{196m}$Au (IT)             &      \\
\hline                                                                                                                                
$^{195m}$Hg&41.6 (8) h    &IT 54.2 (20)          &                &                &11.3                  &$^{192}$Pt($\alpha,n$)       &\cite{A195}\\
\cline{3-5}                                                                                                                          
           &              &EC/$\beta^+$ 45.8 (20)&207.1           &0.37 (8)        &                      &                             &      \\
           &              &                      &261.8           &31 (3)          &                      &                             &      \\
           &              &                      &279.3           &0.14 (4)        &                      &                             &      \\
           &              &                      &368.6           &0.34 (3)        &                      &                             &      \\
           &              &                      &\multicolumn{1}{l}{386.4+387.9}    &2.46 (18)       &                      &                             &      \\
           &              &                      &560.3           &7.1 (5)         &                      &                             &      \\
\cline{1-7}                                                                                                                          
$^{195g}$Hg&10.53 (3) h   &EC/$\beta^+$ 100     &180.1           &1.95 (24)       &11.1                  &$^{192}$Pt($\alpha,n$)       &      \\
           &              &                     &207.1           &1.6 (3)         &                      &$^{195m}$Hg (IT)             &      \\
           &              &                     &261.8           &1.6 (3)         &                      &                             &      \\
           &              &                     &585.1           &2.0 (2)         &                      &                             &      \\
           &              &                     &599.7           &1.83 (22)       &                      &                             &      \\
           &              &                     &779.8           &7.0 (8)         &                      &                             &      \\
           &              &                     &1111.0          &1.48 (22)       &                      &                             &      \\
\cline{1-7}                                                                                                                          
$^{195}$Au &186.01 (6) d  &EC 100               &98.9            &11.21 (15)      &8.8                   &$^{192}$Pt($\alpha,p$)       &      \\
           &              &                     &                &                &                      &$^{195m}$Hg(EC/$\beta^+$)    &      \\
           &              &                     &                &                &                      &$^{195g}$Hg(EC/$\beta^+$)    &      \\
\cline{1-7}                                                                                                                          
$^{195m}$Pt&4.010 (5) d   &IT 100               &98.9            &11.7 (8)        &0.3                   &$^{195}$Pt($\alpha,\alpha'$)  &      \\
\cline{1-7}                                                                                                                          
$^{195m}$Ir&3.67 (8) h    &IT 5 (5)             &                &                &8.5                   &$^{196}$Pt($\alpha,p+\alpha$)&      \\
\cline{3-5}                                                                                                                          
           &              &$\beta^-$ 95 (5)     &98.9            &10.7 (17)       &                      &                             &      \\
\cline{1-7}                                                                                                                          
$^{195g}$Ir&2.29 (17) h   &$\beta^-$ 100        &98.8            &10 (3)          &8.4                   &$^{196}$Pt($\alpha,p+\alpha$)&      \\
           &              &                     &                &                &                      &$^{195m}$Ir (IT)             &      \\
\hline                                                                                                                               
$^{194}$Au &38.02 (10) h  &EC/$\beta^+$ 100     &328.5           &62.8 (16)       &15.1                  &$^{192}$Pt($\alpha,d$)       &\cite{A194}\\
\hline                                                                                                                               
$^{192}$Hg &4.85 (20) h   &EC 100               &99.4            &0.68 (18)       &19.4                  &$^{190}$Pt($\alpha,2n$)      &\cite{A192}\\
           &              &                     &157.2           &7.2 (4)         &                      &                             &      \\
           &              &                     &204.6           &0.85 (18)       &                      &                             &      \\
           &              &                     &262.6           &0.68 (22)       &                      &                             &      \\
           &              &                     &274.8           &52 (4)          &                      &                             &      \\
           &              &                     &279.2           &0.43 (6)        &                      &                             &      \\
\hline                                                                                                                               
$^{51}$Cr  &27.704 (4) d  &EC 100               &320.1           &9.91 (1)        &0.0                   &$^{47}$Ti($\alpha,\gamma$)   &\cite{A51} \\
\hline
\hline
\end{tabular}
\end{center}
\end{table}

\begin{figure}[hbtp]
\begin{center}
\includegraphics[width=1.0\textwidth]{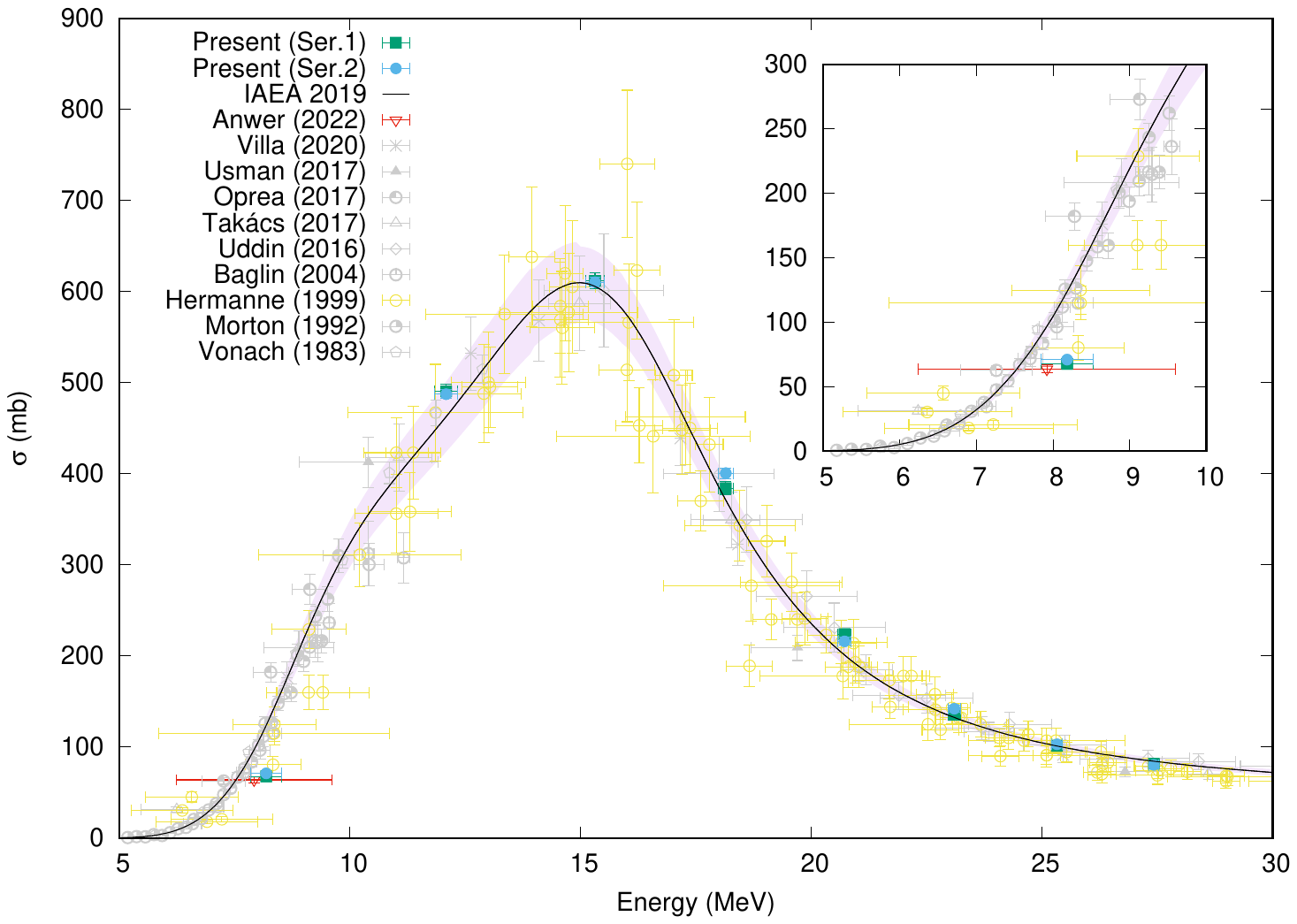}
\end{center}
\caption{
\natu Ti($\alpha$,x)$^{51}$Cr reaction cross sections measured in this work (Series 1 and 2) and recommended by the IAEA~\cite{Hermanne2018} compared with the literature data not rejected in preparation of the IAEA recommendation~\cite{Anwer2022,Villa2020,Usman2017,Oprea2017,Takacs2017,Uddin2016,Baglin2005,Hermanne1999,Morton1992,Vonach1983}.
The error bars of the cross sections in the present work are for the uncertainties propagated from those in the peak areas.
The shaded area accompanying the line indicates the uncertainty in the cross sections recommended by the IAEA.
}
\label{fig:Ti_a_x_51Cr}
\end{figure}

Figure~\ref{fig:Ti_a_x_51Cr} shows comparison of the \natu Ti($\alpha$,x)$^{51}$Cr reaction cross sections measured in the present work with those recommended by the IAEA~\cite{Hermanne2018} and in the literature~\cite{Anwer2022,Villa2020,Usman2017,Oprea2017,Takacs2017,Uddin2016,Baglin2005,Hermanne1999,Morton1992,Vonach1983}.
We determined the beam energy at each foil and beam flux by using the SRIM code and integrated charge, respectively.
The measured cross sections are consistent with the recommended cross sections except for the lowest energy point at 8.2~MeV,
where the energy spread is relatively high in the present measurement.
We adopted the measured beam flux and target thicknesses without any adjustment.

\section{Data analysis using the activation cross section formula}
As the first step of the analysis of the measured $\gamma$ spectra, we studied cooling time dependence of the isomeric ratios ($\sigma_g/\sigma_m$) of $^{198}$Au, $^{197}$Hg and $^{195}$Hg for the first Pt foil (29~MeV) by using Eq.~(\ref{eqn:cumgm1}).
The analyzed $\gamma$ lines are at
\begin{itemize}
\item 204 and 215~keV for $^{198m}$Au, and 412~keV for $^{198g}$Au
\item 134 and 279~keV for $^{197m}$Hg, and 191~keV for $^{197g}$Hg
\item 261 and 560~keV for $^{195m}$Hg, and 780~keV for $^{195g}$Hg.
\end{itemize}
Among these lines,
\begin{itemize}
\item 279~keV $\gamma$ (6.1\%) of $^{197m}$Hg is also a characteristic $\gamma$ line of $^{197m}$Pt (2.4\%) and $^{195m}$Hg (0.14\%)
\item 191~keV $\gamma$ (0.632\%) of $^{197g}$Hg is also a characteristic $\gamma$ line of $^{197g}$Pt (3.7\%)
\item 261~keV $\gamma$ (31\%) of $^{195m}$Hg is also a characteristic $\gamma$ line of $^{195g}$Hg (1.6\%).
\end{itemize}
The contributions of these interfering $\gamma$ lines were not subtracted at this stage.
Note that we did not find a characteristic peak of $^{197g}$Pt usable for quantification of its production cross section in the present work.

\begin{figure}[hbtp]
\begin{center}
\includegraphics[width=0.8\textwidth]{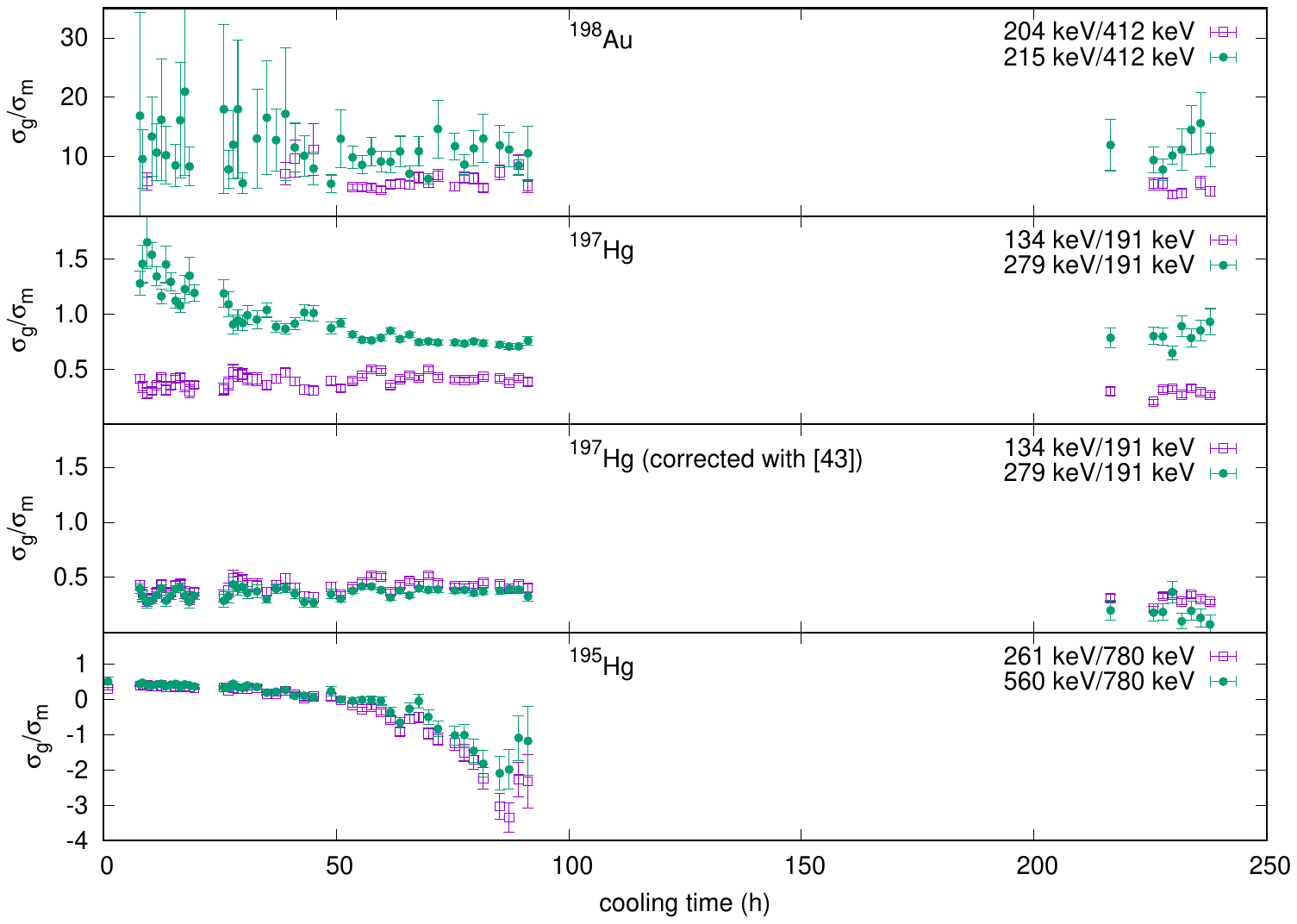}
\end{center}
\caption{
Cooling time dependence of the isomeric ratios ($\sigma_g/\sigma_m$) of $^{198}$Au, $^{197}$Hg and $^{195}$Hg for the irradiation at 29~MeV by applying Eq. (1) to several characteristic $\gamma$ lines.
The $\gamma$ energies before and after the slash in the figure legends are for the $\gamma$ lines adopted for quantification of the metastable state and ground state, respectively.
The decay parameters in the ENSDF library~\cite{A198,A197,A195} were used to determine the isomeric ratios.
For $^{197}$Hg,
the isomeric ratios derived with the IT branching ratio and $\gamma$ emission probabilities published by Lebeda et al.~\cite{Lebeda2019} are additionally shown.
The error bars are for the uncertainties propagated from those in the peak areas.
}
\label{fig:tcdep}
\end{figure}

Figure~\ref{fig:tcdep} shows the cooling time dependence of the isomeric ratios.
We do not see strong cooling time dependence in the isomeric ratios of $^{198}$Au and the isomeric ratios of $^{197}$Hg obtained from the 134 and 191~keV $\gamma$ counts.
On the other hand, we observe strong cooling time dependence in the isomeric ratio of $^{197}$Hg obtained from the 279 and 191~keV $\gamma$ counts and the isomeric ratios of $^{195}$Hg. 

Lebeda et al. studied the $\gamma$ emission probabilities of $^{197m}$Hg~\cite{Lebeda2020}.
Their work is motivated by inconsistency in the $^{197}$Au($p,n$)$^{197m}$Hg and $^{197}$Au($d,2n$)$^{197m}$Hg reaction cross sections measured by them with the 134 and 279~keV $\gamma$ lines~\cite{Cervenak2019,Lebeda2019}.
They performed an experiment dedicated to determination of the $^{197m}$Hg isomeric transition (IT) branching ratio, and suggested revision of the emission probabilities in the ENSDF library~\cite{A197} from 33.5\% to 34.8\% for the 134~keV $\gamma$ line and from 6.1\% to 3.79\% for the 279~keV $\gamma$ line.
The revision is minor for the 134~keV $\gamma$ emission probability but significant for the 279~keV $\gamma$ emission probability.
Figure~\ref{fig:tcdep} shows that the cooling time dependence of the isomeric ratio determined with the 279~keV $\gamma$ line is resolved and is very close to the isomeric ratio determined with the 134~keV $\gamma$ line if we adopt the IT branching ratio (94.68$\pm$0.09\%) and $\gamma$ emission probabilities determined by Lebeda et al. instead of those in the ENSDF library.
Elimination of the cooling time dependence by updating the IT branching ratio and $\gamma$ emission probabilities for $^{197m}$Hg would suggest that a similar problem may exist in the $^{195m}$Hg IT branching ratio and $\gamma$ emission probabilities in the ENSDF library~\cite{A195}.

\section{Data analysis using simultaneous decay curve analysis}
To investigate the cooling time dependence of the isomeric ratios further,
we performed least-squares analysis for the peak areas of all characteristic $\gamma$ lines with the production cross sections of 16 nuclides and IT branching ratios of $^{197m}$Hg and $^{195m}$Hg as adjustable parameters.
We refer this approach as simultaneous decay curve analysis.
We will see that the simultaneous decay curve analysis allows us to utilize a $\gamma$ line attributed to several $\gamma$ emitting nuclides having similar $\gamma$ energies or sharing the common daughter nuclide.
Such $\gamma$ lines are often excluded in analysis for conventional activation cross section determinations.

Let us denote an efficiency corrected $\gamma$ emission rate of $\gamma$ line $s$ measured at cooling time $t_c$ by $A_s(t_c)$,
and consider that the $\gamma$ line may be emitted from $I$ different nuclides, namely,
\begin{equation}
A_s(t_c)=\sum_{i=1}^I I_{si}\lambda_iN_i(t_c),
\end{equation}
where $N_i(t_c)$ is the number of atoms of the nuclide $i$ at cooling time $t_c$,
and $I_{si}$ is the emission probability of the $\gamma$ line $s$ following decay of the nuclide $i$ with the decay constant $\lambda_i$.

Formation of the nuclide $i$ may be expressed by superposition of the processes producing a nuclide $j$ ($j=1,i$) followed by a series of decays $j$ $\to$ $j+1$ $\to$ $j+2$ $\to$ $\cdots$ $\to$ $i$, namely,
\begin{equation}
N_i(t_c)=\phi n \sum_{j=1}^i \sigma_j p_{ji} 
                \sum_{k=j}^i \Lambda_k^{j,i}
         \frac{(1-e^{-\lambda_k t_b})e^{-\lambda_k t_c}}{\lambda_k}
\label{eqn:solutiongen}
\end{equation}
with the branching ratio $p_{ji}$ for the decay chain from $j$ to $i$ ($p_{ii}=1$), beam flux $\phi$, and sample atom number areal density $n$.\footnote{
See Appendix~\ref{sec:app5} for its proof.
}
In Eq.~(\ref{eqn:solutiongen}),
the first summation is taken over all reaction products $j$ decaying toward the nuclide $i$ including the nuclide $i$ itself ($j=i$),
while the second summation is taken over all nuclides $k$ on the decay chain from $j$ to $i$.
By defining the production rate $R_i=\phi n \sigma_i$ and 
\begin{equation}
g_k(t_c)=\frac{(1-e^{-\lambda_k t_b})e^{-\lambda_k t_c}}{\lambda_k},
\end{equation}
Eq.~(\ref{eqn:solutiongen}) is simplified to
\begin{equation}
N_i(t_c)=\sum_{j=1}^i R_j p_{ji} 
         \sum_{k=j}^i \Lambda_k^{j,i}g_k(t_c).
\label{eqn:solution}
\end{equation}
Hereafter, we omit indication of $t_c$ dependence of $N_i$ and $g_k$.

If the nuclide \textit{1} does not have any decay precursor, Eq.~(\ref{eqn:solution}) gives the number of atoms of the nuclide \textit{1} after cooling for $t_c$ as
\begin{equation}
N_1=R_1g_1.
\label{eqn:solution1}
\end{equation}
If the nuclide \textit{1} decays into the nuclide \textit{2} with the branching ratio $p_{12}$,
and the nuclide \textit{1} is the unique precursor of the nuclide \textit{2},
Eq.~(\ref{eqn:solution}) gives the number of atoms of the nuclide \textit{2} after cooling for $t_c$ as
\begin{equation}
N_2=R_1p_{12}\left[
 \frac{\lambda_1}{\lambda_2-\lambda_1}g_1
+\frac{\lambda_1}{\lambda_1-\lambda_2}g_2
\right]
+R_2g_2,
\label{eqn:solution2}
\end{equation}
where the first term describes production of the nuclide \textit{2} due to decay of the nuclide \textit{1},
while the second term is for direct production of the nuclide \textit{2}.
Similarly,
the number of atoms of the nuclide \textit{3} after cooling for $t_c$ is
\begin{equation}
N_3 = R_1 p_{13} \left[
\frac{\lambda_1\lambda_2}{(\lambda_2-\lambda_1)(\lambda_3-\lambda_1)}g_1
+
\frac{\lambda_1\lambda_2}{(\lambda_1-\lambda_2)(\lambda_3-\lambda_2)}g_2
+
\frac{\lambda_1\lambda_2}{(\lambda_1-\lambda_3)(\lambda_2-\lambda_3)}g_3
\right]
+
R_2p_{23}\left[
 \frac{\lambda_2}{\lambda_3-\lambda_2}g_2
+\frac{\lambda_2}{\lambda_2-\lambda_3}g_3
\right]
+R_3g_3
\label{eqn:solution3}
\end{equation}
if the nuclide \textit{2} is the unique decay precursor of the nuclide \textit{3}.

Table~\ref{tab:decayterm} summarizes the decay chains of the 27 $\gamma$ lines modelled in the present simultaneous decay curve analysis.
For example,
this table shows that the 261~keV $\gamma$ line originates from the decay of $^{195m}$Hg, $^{195g}$Hg and $^{192}$Hg following their production:
\begin{itemize}
\item \natu Pt($\alpha$,x)$^{195m}$Hg (EC/$\beta^+$)
\item \natu Pt($\alpha$,x)$^{195m}$Hg (IT)$^{195g}$Hg (EC/$\beta^+$)
\item \natu Pt($\alpha$,x)$^{195g}$Hg (EC/$\beta^+$)
\item \natu Pt($\alpha$,x)$^{192}$Hg (EC).
\end{itemize}
The $^{192}$Hg $\gamma$ line (262.6~keV) is located about 0.8~keV higher than the $^{195}$Hg $\gamma$ line (261.8~keV),
but they are treated as unresolved in our analysis.
Among the three directly produced nuclides,
$^{195m}$Hg contributes to the 261~keV $\gamma$ emission by its one- or two-step decay (``1" and ``2" in Table~\ref{tab:decayterm}),
while $^{195g}$Hg and $^{192}$Hg contribute by one-step decay.
By applying Eqs.~(\ref{eqn:solution1}) and (\ref{eqn:solution2}) to the four decay processes emitting 261~keV $\gamma$-rays,
its emission rate $A_{261}$ is expressed by
\begin{equation}
A_{261}=I_m\lambda_mR_mp_\beta g_m
       +I_g\lambda_gR_mp_\mathrm{IT}\left[
        \frac{\lambda_m}{\lambda_g-\lambda_m}g_m
       +\frac{\lambda_m}{\lambda_m-\lambda_g}g_g
       \right]
       +I_g\lambda_g R_g g_g
       +I_2\lambda_2 R_2 g_2,
\label{eqn:A261}
\end{equation}
where the suffices $m$, $g$ and 2 refer into $^{195m}$Hg, $^{195g}$Hg and $^{192}$Hg.
The branching ratios of $^{195m}$Hg by EC/$\beta^+$ and IT decay are denoted by $p_\beta$ and $p_\mathrm{IT}$,
respectively.
(The decay branching ratios of $^{195g}$Hg and $^{192}$Hg are 1.)

\begin{landscape}
\begin{table}[hbtp]
\caption{
Contribution of reaction product nuclides to $\gamma$ lines modelled in the present work. ``1", ``2" and ``3" mean that the reaction product nuclide emits the $\gamma$-ray following one $\beta$/EC/IT decay process (e.g., 98~keV $\gamma$ emission following \natu Pt($\alpha$,x)$^{195}$Au EC decay),
two $\beta$/EC/IT processes (e.g., 98~keV $\gamma$ emission following \natu Pt($\alpha$,x)$^{195g}$Hg EC/$\beta^+$ decay to $^{195}$Au and its EC decay),
and three $\beta$/EC/IT processes (e.g., 98~keV $\gamma$ emission following \natu Pt($\alpha$,x)$^{195m}$Hg IT decay to $^{195g}$Hg,
its EC/$\beta^+$ decay to $^{195}$Au and its EC decay), respectively.
}
\label{tab:decayterm}
\begin{center}
\begin{tabular}{rcccccccccccccccc}
\hline
\hline
$E_\gamma$ (keV)&$^{200m}$Au&$^{200g}$Au&$^{199m}$Hg&$^{199}$Au&$^{198m}$Au&$^{198g}$Au&$^{197m}$Hg&$^{197g}$Hg&$^{196m}$Au&$^{196g}$Au&$^{195m}$Hg&$^{195g}$Hg&$^{195}$Au&$^{195m}$Pt&$^{194}$Au&$^{192}$Hg\\
\hline
    98          &           &           &           &          &           &           &           &           &           &           &2,3        &2          &1         &1          &          &1         \\
   133          &1          &           &           &          &           &           &1          &           &           &           &           &           &          &           &          &          \\
   147          &           &           &           &          &           &           &           &           &1          &           &           &           &          &           &          &          \\
   158          &           &           &1          &1         &           &           &           &           &           &           &           &           &          &           &          &1         \\
   180          &           &           &           &          &1          &           &           &           &           &           &2          &1          &          &           &          &          \\
\hline
   188          &           &           &           &          &           &           &           &           &1          &           &           &           &          &           &          &          \\
   191          &           &           &           &          &           &           &2          &1          &           &           &           &           &          &           &          &          \\
   204          &           &           &           &          &1          &           &           &           &           &           &           &           &          &           &          &1         \\
   208          &           &           &           &1         &           &           &           &           &           &           &1,2        &1          &          &           &          &          \\
   214          &           &           &           &          &1          &           &           &           &           &           &           &           &          &           &          &          \\
\hline
   255          &1          &           &           &          &           &           &           &           &           &           &           &           &          &           &          &          \\
   261          &           &           &           &          &           &           &           &           &           &           &1,2        &1          &          &           &          &1         \\
   274          &           &           &           &          &           &           &           &           &           &           &           &           &          &           &          &1         \\
   279          &           &           &           &          &           &           &1          &           &           &           &1          &           &          &           &          &1         \\
   328          &           &           &           &          &           &           &           &           &           &           &           &           &          &           &1         &          \\
\hline
   333          &1          &           &           &          &1          &           &           &           &2          &1          &           &           &          &           &          &          \\
   355          &           &           &           &          &           &           &           &           &2          &1          &           &           &          &           &          &          \\
   367          &1,2        &1          &           &          &           &           &           &           &           &           &1          &           &          &           &          &          \\
   374          &           &           &1          &          &           &           &           &           &           &           &           &           &          &           &          &          \\
   387          &           &           &           &          &           &           &           &           &           &           &1          &           &          &           &          &          \\
\hline
   411          &           &           &           &          &2          &1          &           &           &           &           &           &           &          &           &          &          \\
   560          &           &           &           &          &           &           &           &           &           &           &1          &           &          &           &          &          \\
   579          &1          &           &           &          &           &           &           &           &           &           &           &           &          &           &          &          \\
   585          &           &           &           &          &           &           &           &           &           &           &2          &1          &          &           &          &          \\
   599          &           &           &           &          &           &           &           &           &           &           &2          &1          &          &           &          &          \\
\hline
   779          &           &           &           &          &           &           &           &           &           &           &2          &1          &          &           &          &          \\
  1111          &           &           &           &          &           &           &           &           &           &           &2          &1          &          &           &          &          \\
\hline
\hline
\end{tabular}
\end{center}
\end{table}
\end{landscape}

The above-mentioned formulation shows that the $\gamma$ emission rate can be expressed as a linear combination of the production cross sections.
For example, Eq.~(\ref{eqn:A261}) can be rewritten as a linear combination of the production cross sections of $^{195m}$Hg, $^{195g}$Hg and $^{192}$Hg.
If we denote a set of the emission rates of various $\gamma$ lines and a set of the production cross sections of various nuclides by column vectors $\bm{A}=\{A_i\}$ and $\bm{\sigma}=\{\sigma_i\}$,
respectively,
the relationship between the two vectors at a given incident energy may be modelled by the observation equation
\begin{equation}
\bm{A}=G\bm{\sigma}+\bm{e}
\end{equation}
with the design matrix $G$ and a vector $\bm{e}$ expressing the residual (fitting deviation).
As Table~\ref{tab:decayterm} is for 16 product nuclides and 27 $\gamma$ lines,
the vectors $\bm{\sigma}$ and $\bm{A}$ have 16 and 27 elements,
respectively,
and $G$ is a 16$\times$27 matrix.
The least-squares solution of the observation equation gives the production cross sections and their covariance $M$:
\begin{eqnarray}
\bm{\sigma}&=&(G^TV^{-1}G)^{-1}G^TV^{-1}\bm{A}\\
M&=&(G^TV^{-1}G)^{-1},
\label{eqn:GLSM}
\end{eqnarray}
where $V$ is a diagonal matrix with the variances of the measured $\gamma$ emission rates as the diagonal elements,
which originates from the uncertainty in the peak area in the present work.

In the present work, we included the IT branching ratios of $^{197m}$Hg and $^{195m}$Hg as additional parameters.
Since the IT branching ratio and metastable state production cross section appear as their product in the observation equation,
adjustment of the IT branching ratios and production cross sections together based on the observation equation is no longer a linear least-squares problem.
In the present work, we adopted the Marquardt-Levenberg algorithm~\cite{Levenberg1944,Marquardt1963} implemented in the computer program \textsc{gnuplot}~\cite{gnuplot52} to solve the equation including the IT branching ratios as fitting parameters.
The branching ratios were determined by fitting for the first Pt foil irradiated at 29~MeV because many data points are available for this foil thanks to the repeated measurements, and the branching ratios adjusted at 29~MeV were treated as constants in the subsequent analysis of the Pt foils bombarded at lower energies.

Among the nuclides in Table~\ref{tab:decaydata},
we excluded $^{197}$Pt from the least-squares analysis at all energies since
(1) the $^{197m}$Pt 346.5~keV peak area is seen only in one measurement at each of three Pt foils, and
(2) the $^{197g}$Pt production cross sections become negative if we include them in fitting.
Similarly, we set the production cross section to zero for a Pt foil if
(1) the most intense $\gamma$ line of the nuclide is absent in any spectrum, or
(2) fitting procedure makes the production cross section negative.

Our preliminary decay curve analysis of the 98~keV $\gamma$ emission rate at 29~MeV including $^{195m}$Ir and $^{195g}$Ir contributions showed absence of the sensitivity of the measured $\gamma$ emission rate to the production cross sections of these nuclides (i.e., initial guesses became adjusted values but with huge uncertainties), and we excluded them from the fitting procedure for all Pt foils as well.
Note that none of the other intense $\gamma$ lines of $^{195m}$Ir and $^{195g}$Ir were observed in the spectra at 29~MeV. 
Additionally, we excluded the $^{195}$Au and $^{195m}$Pt production cross sections from fitting at 24.2 and 21.9~MeV since these two cross sections can be determined only from the 98~keV $\gamma$ emission rate,
and only one measurement detected its peak area at these incident energies (i.e., unsolvable). 

\section{Results}
\subsection{Decay curves}
\begin{figure}[hbtp]
\begin{center}
\includegraphics[width=1.0\textwidth]{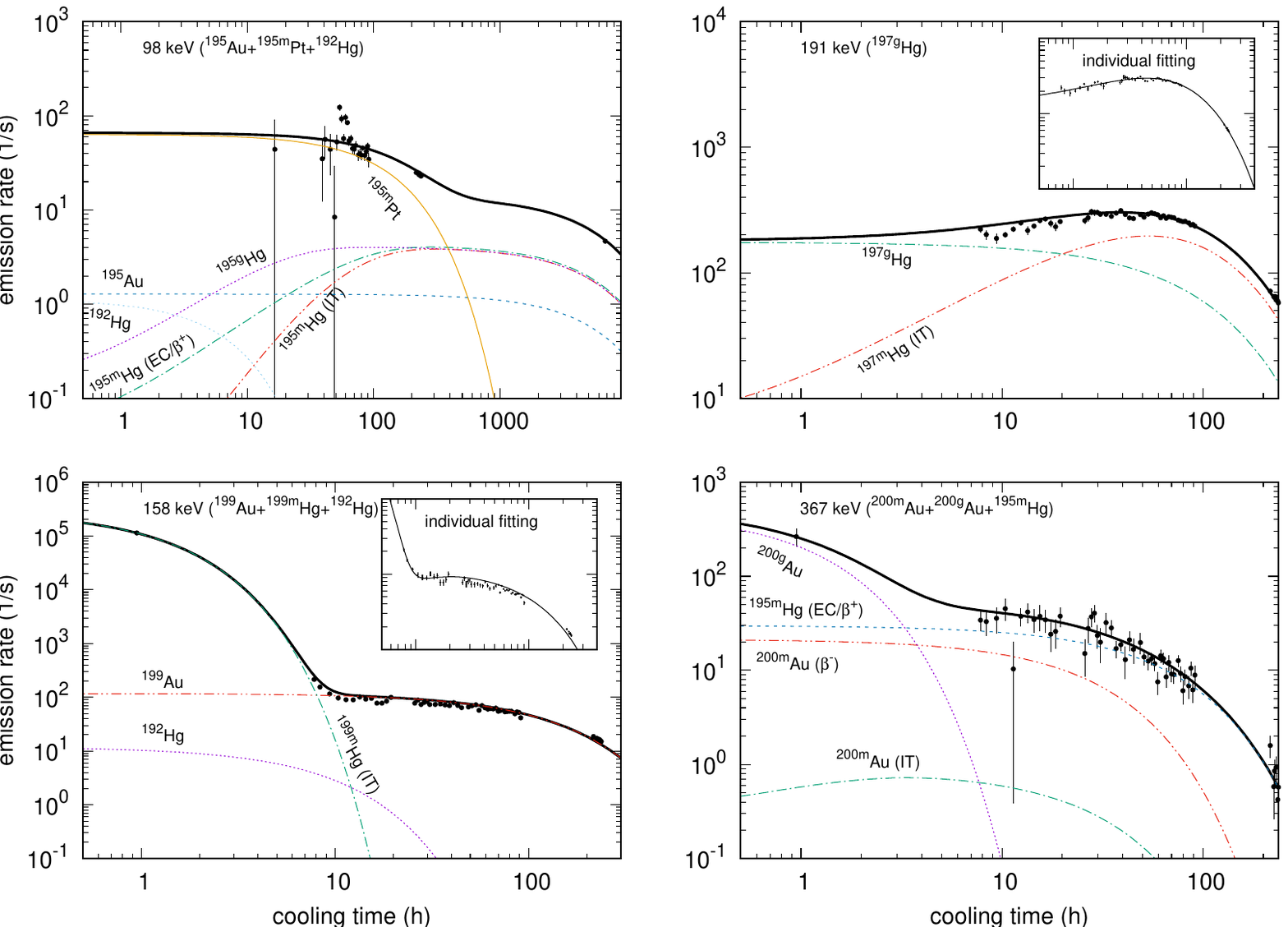}
\end{center}
\caption{
Emission rates of four $\gamma$ lines and their decompositions to the contributions from various reaction products by simultaneous decay curve analysis for the Pt foil irradiated at 29~MeV.
The parenthesized nuclide symbols indicate the nuclides emitting the $\gamma$ line,
while the nuclide symbol near each partial decay curve explains the reaction product contributing to the emission rate of the $\gamma$ line.
The error bars of the measured rates are for the uncertainties propagated from those in the peak areas.
The inset gives the individual fitting to the emission rate of the particular $\gamma$ line.
}
\label{fig:decaycurve}
\end{figure}

Figure~\ref{fig:decaycurve} shows the measured $\gamma$ emission rates corrected for the detection efficiencies along with the total and partial decay curves adjusted to reproduce the measured emission rates for four selected $\gamma$ lines.
The parenthesized nuclide symbols show the nuclides emitting the $\gamma$-rays,
which may be one directly produced by a reaction or by decay of another reaction product.
The nuclide symbol on each partial decay curve explains the reaction product contributing to the emission rate shown by the curve.
The decay branch (e.g., IT) is indicated for the $\gamma$ lines originating from the production of a metastable state since it may have two branches contributing to the same $\gamma$ line (e.g., EC/$\beta^+$ decay of the ground state following the IT decay of the metastable state, or EC/$\beta^+$ decay of the same metastable state).

Among the four $\gamma$ lines,
the emission rates were detectable for the 158 and 367~keV $\gamma$ lines in the first $\gamma$ spectrum taken about 1~h after the end of bombardment because of two short-lived products $^{199m}$Hg and $^{200g}$Au.
The figure shows that the first measurement was essential to quantify their production cross sections.
The first measurement also contributed to obtain the correct time profile of the decay curves at the early stage of cooling because the second measurement was performed at about 8 h after the end of bombardment.

Table~\ref{tab:decayterm} shows that the 98 keV $\gamma$ emission rate is only one that requires consideration of decays through four generations ($^{195m}$Hg → $^{195g}$Hg → $^{195}$Au → $^{195g}$Pt).
Contribution of the $^{192}$Hg production is minor since its 98 keV $\gamma$ emission probability is low and its production cross section is also very low (less than 0.1~mb).
%
%
Cooling time dependence of the partial decay curves originating from the direct production of $^{195}$Au and $^{195m}$Pt are expressed by single exponential curves.
These two nuclides have similar production cross sections ($\sigma\sim$3~mb) and emission probabilities ($\sim$11\%), but their decay curves are different on the time scale because of the large difference in their half-lives.
The $^{195m}$Hg and $^{195g}$Hg nuclei directly produced by reactions gradually decay to $^{195}$Au.
The four partial decay curves originating from the productions of $^{195m}$Hg, $^{195g}$Hg and $^{195}$Au start to show the same cooling time dependence after complete decay of $^{195m}$Hg and $^{195g}$Hg to $^{195}$Au.
It will be shown in Sect.~\ref{sec:excfun} that the three cross sections corresponding to the $^{195m}$Hg (IT), $^{195m}$Hg (EC/$\beta^+$) and $^{195g}$Hg productions are almost equal ($\sim$10~mb),
and this explains agreement of the three partial decay curves after long cooling.

Our simultaneous decay curve analysis becomes the usual one if we perform fitting to the emission rate of a particular $\gamma$ line separately.
Individual fitting sometimes improves fitting,
but it is not always the case.
To demonstrate this point,
we added the individual fitting results in the insets for the 158 and 191~keV $\gamma$ line cases.
The reduced chi-square ($\chi^2$ per degree of freedom, $\chi^2_\mathrm{red}$) from the simultaneous least-squares fitting for all 27 $\gamma$ lines (5.77) is improved to 1.64 for individual fitting for the 191~keV $\gamma$ line case,
and we observe an improvement in the individual fitting result at the early stage of cooling in the inset ($t_c\sim$10~h).
Contrary,
individual fitting increases $\chi^2$ from 5.77 to 9.55 for the 158~keV $\gamma$ line case.
The inset shows that the individual fitting improves agreement at $t_c\sim$10~h,
but does not improve the residual seen between $t_c\sim$ 20 to 100~h.

\subsection{Cooling time dependence of $^{198}$Au, $^{197}$Hg and $^{195}$Hg isomeric ratios at 29~MeV}
\begin{figure}[hbtp]
\begin{center}
\includegraphics[width=1.0\textwidth]{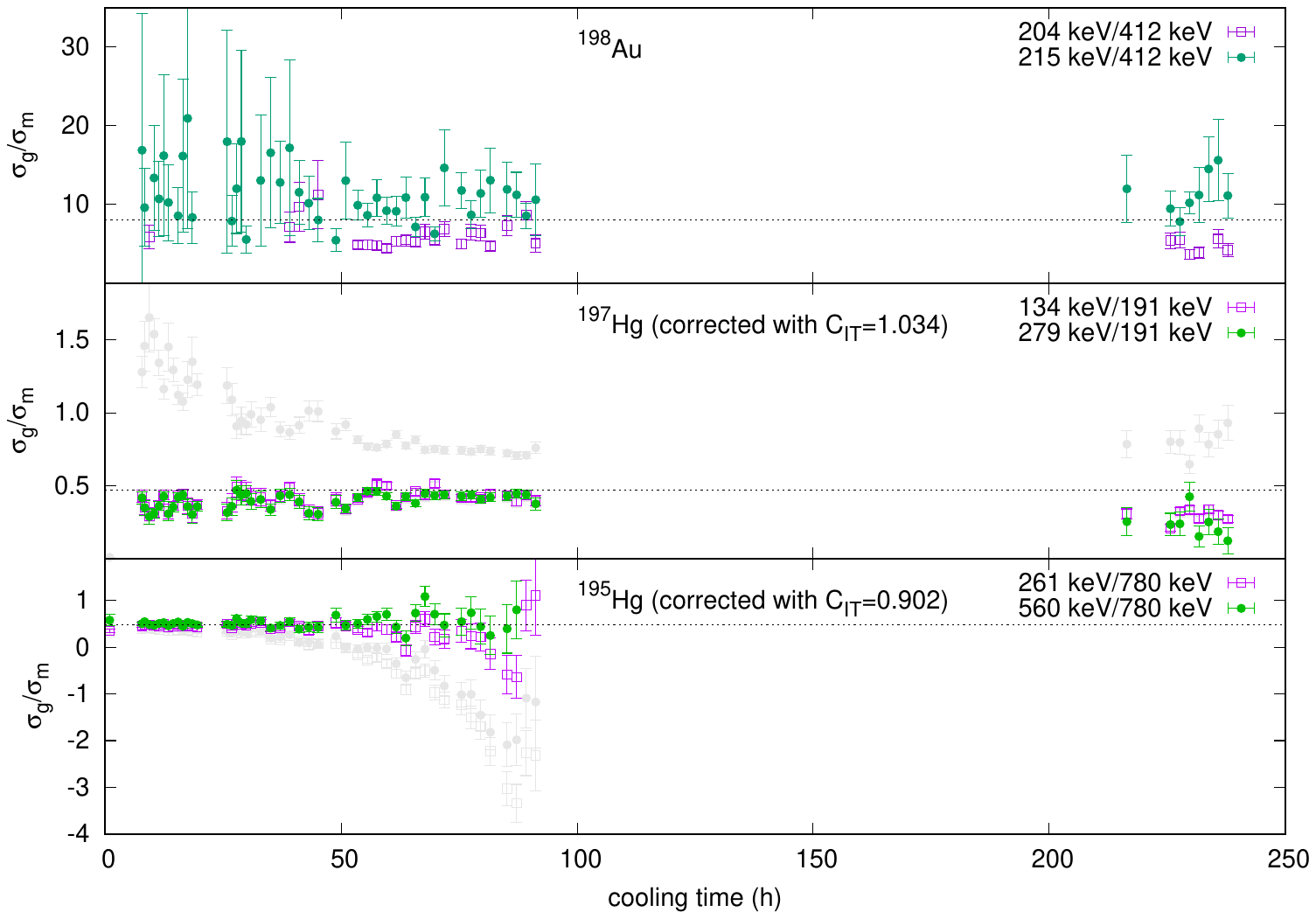}
\end{center}
\caption{
Cooling time dependence of the isomeric ratios of $^{198}$Au, $^{197}$Hg and $^{195}$Hg determined for the Pt foil irradiated at 29~MeV by applying Eq.~(\ref{eqn:cumgm1}) to several characteristic $\gamma$ lines.
For $^{197}$Hg and $^{195}$Hg,
the isomeric ratios derived with the corrected decay data (Table~\ref{tab:decaydatarev}) are plotted along with those derived with the ENSDF decay data in gray symbols.
The horizontal dashed line shows the isomeric ratio derived from the simultaneous decay curve analysis.
See also the caption of Fig.~\ref{fig:tcdep}.
}
\label{fig:tcdeprev}
\end{figure}

The least-squares analysis at 29~MeV gives the correction factors (multiplication factors applied to the ENSDF IT branching ratios) $C_\mathrm{IT}=$1.03356$\pm$0.00016 for $^{197m}$Hg and 0.9020$\pm$0.0064 for $^{195m}$Hg.
Accordingly,
we corrected the branching ratios and $\gamma$ emission probabilities in Table~\ref{tab:decaydata} as summarized in Table~\ref{tab:decaydatarev},
and adopted the corrected ones in the subsequent decay curve analysis.
This table shows that the $^{195m}$Hg branching ratios corrected by us are very close to those measured by Lebeda et al.~\cite{Lebeda2020}.
Figure~\ref{fig:tcdeprev} shows the cooling time dependence of the isomeric ratios of these nuclides with and without considering the correction factors.
It shows that the cooling time dependence seen in Fig.~\ref{fig:tcdep} is successfully eliminated by adoption of the corrected IT branching ratios and $\gamma$ emission probabilities. 

\begin{table}[hbtp]
\caption{
Decay branching ratios and $\gamma$ emission probabilities in the ENSDF library and those corrected in the present work.
The $I_\gamma$ values and their uncertainties are those derived by the IAEA LiveChart of Nuclides~\cite{Verpelli2011} from the ENSDF library.
All corrected values were determined by applying the correction factor $C_\mathrm{IT}$ to the ENSDF values.
See the Table~\ref{tab:decaydata} caption for the meaning of the parenthesized numbers.
}
\label{tab:decaydatarev}
\begin{center}
\begin{tabular}{llllllllll}
\hline
\hline
           &$C_\mathrm{IT}$  &Mode        &\multicolumn{3}{c}{$p$ (\%)}                          &$E_\gamma$ (keV)&\multicolumn{3}{c}{$I_\gamma$ (\%)}           \\
\cline{4-6}                
\cline{8-10}                
            &                &            &ENSDF\cite{A197,A195}     &Corrected&Lebeda~\cite{Lebeda2020}&                &ENSDF\cite{A197,A195}&Corrected &Lebeda~\cite{Lebeda2020}\\
\hline                                                                                                
$^{197m}$Hg &1.03356 (16)    &IT          &91.4 (7)                  &94.5 (7) &94.68 (9)               &134.0           &33.5 (3)             &34.6 (3)  &34.8 (3)                \\
\cline{3-10}                                                                                                                                                                     
            &                &EC          &8.6 (7)                   &5.5 (7)  &5.32 (9)                &279.0           &6.1 (5)              &3.9 (7)   &3.79 (4)                \\
\hline                                                                                                                                                                          
$^{195m}$Hg &0.9020 (64)     &IT          &54.2 (20)                 &48.9 (18)&                        &                &                     &          &                        \\
\cline{3-10}                                                                                                                                                                     
            &                &EC/$\beta^+$&45.8 (20)                 &51.1 (18)&                        &207.1           &0.37 (8)             &0.41 (9)  &                        \\
            &                &            &                          &         &                        &261.8           &31 (3)               &35 (4)    &                        \\
            &                &            &                          &         &                        &279.3           &0.14 (4)             &0.16 (5)  &                        \\
            &                &            &                          &         &                        &368.6           &0.34 (3)             &0.38 (4)  &                        \\
            &                &            &                          &         &                        &386.4+387.9     &2.46 (18)            &2.75 (25) &                        \\
            &                &            &                          &         &                        &560.3           &7.1 (5)              &7.9 (7)   &                        \\
\hline
\hline
\end{tabular}
\end{center}
\end{table}

The current ENSDF evaluation~\cite{A195} adopts the $^{195m}$Hg IT branching ratio of 54.2$\pm$2.0\% based on the experimental data published in~\cite{Vieu1973}.
However,
the ENSDF evaluators also report the presence of other experiments suggesting lower IT branching ratios such as 49\%~\cite{Canty1970}, 47.8$\pm$0.5\%~\cite{Frank1967} and 51\%~\cite{Frana1971}.
The isomeric transition of $^{195m}$Hg does not emit a $\gamma$-ray suitable for the usual $\gamma$ spectrometry,
hence the experimental approach applied to $^{197m}$Hg by Lebeda et al. would be inapplicable to $^{195m}$Hg.
Nevertheless, we would suggest a new measurement of the $^{195m}$Hg IT branching ratio to confirm our finding.
We also plan to investigate if the correction factor determined by us is influenced by $^{195}$Ir production by another experiment covering the higher energy.

\subsection{Excitation functions}
\label{sec:excfun}
Cross sections for production of 18 nuclides and two IT branching ratios were determined at 29 MeV.
Additional cross sections for production of fewer nuclides were also determined at lower energies.
These data were derived by the simultaneous decay curve analysis applied to each Pt foil separately.
The obtained cross sections are \textit{independent},
ignoring the presence of short-lived ($T_{1/2}<$ 1~min) precursors.
The sum of the ground state and metastable state production cross sections $\sigma_{g+m}=\sigma_g+\sigma_m$,
and isomeric ratios $IR=\sigma_m/(\sigma_g+\sigma_m$) were derived from $\sigma_g$ and $\sigma_m$.

Some parameters characterizing the least-squares fitting are summarized in Table~\ref{tab:fitstatistics}.
This table shows that very few experimental data points were available at the three lowest energies.
For example, only two data points of 98~keV $\gamma$-rays were usable for the decay curve analysis at 5.7~MeV, where only the $^{195}$Pt($\alpha$,$\alpha'$)$^{195m}$Pt reaction can contribute to the $\gamma$ counts,
and its production cross section of $^{195m}$Pt was determined.
Tables~\ref{tab:198Au} to \ref{tab:195Au195Pt194Au192Hg} list the production cross sections and isomeric ratios adjusted by the simultaneous decay curve analysis.
The present data are also plotted in Figs.~\ref{fig:198Au} to \ref{fig:192Hg} and compared with the experimental datasets in the literature if they exist.
Capurro et al.~\cite{Capurro1991} published the elemental cross sections divided by the atomic mass of natural platinum (c.f. the editorial note of~\cite{Bonesso2017}),
and they are converted to the elemental cross sections in these figures.
The uncertainties (one standard deviation) of the cross sections and isomeric ratios in the tables and figures correspond to the variances $M$ in Eq.~(\ref{eqn:GLSM}) but multiplied by $\chi^2_\mathrm{red}$ (``external uncertainties"~\cite{Birge1932,Mughabghab2018}) for better consistency between the measured peak areas and adjusted parameters (i.e., $\chi^2_\mathrm{red}=1$).
The mean and difference of the entrance and exit $\alpha$-particle energies of each foil are given under $E$ and $\Delta E$ in the tables and also error bars in the figures.

The excitation functions of the studied reactions were predicted by the statistical and pre-equilibrium models using TALYS-2.0~\cite{Koning2023}.
They are calculated with the spin cutoff parameter defined by $\sigma^2 = \eta I_\mathrm{rig} T/\hbar$ with the rigid body moment of inertia $I_\mathrm{rig}$, nuclear temperature $T$ and spin cutoff parameter $\eta$ (see Ref.~\cite{Rodrigo2023} for its treatment in TALYS).
The model predictions with $\eta=1$ are plotted in the figures unless specified otherwise. 

\begin{table}[hbtp]
\caption{
Degree of freedom (difference between the number of data points and number of fitting parameters), reduced chi-square ($\chi^2$ per degree of freedom, $\chi^2_\mathrm{red}$) and its square root.
}
\label{tab:fitstatistics}
\begin{center}
\begin{tabular}{lrrrrrrrrr}
\hline
\hline
$E$ (MeV)                           &28.4  &26.4  &24.2  &21.9  &19.5  &16.8  &13.7  &10.2  &5.7   \\
\hline
Degree of freedom                   &1021  &25    &21    &18    &27    &8     &1     &1     &1     \\
$\chi^2_\mathrm{red}$               &5.77  &9.21  &9.17  &3.74  &13.2  &5.00  &1.75  &4.49  &0.51  \\
Square root of $\chi^2_\mathrm{red}$&2.40  &3.03  &3.03  &1.93  &3.63  &2.24  &1.32  &2.12  &0.71  \\
\hline
\hline
\end{tabular}
\end{center}
\end{table}

\subsubsection{$^{198}$Au}
\begin{figure}[hbtp]
\begin{center}
\includegraphics[width=0.9\textwidth]{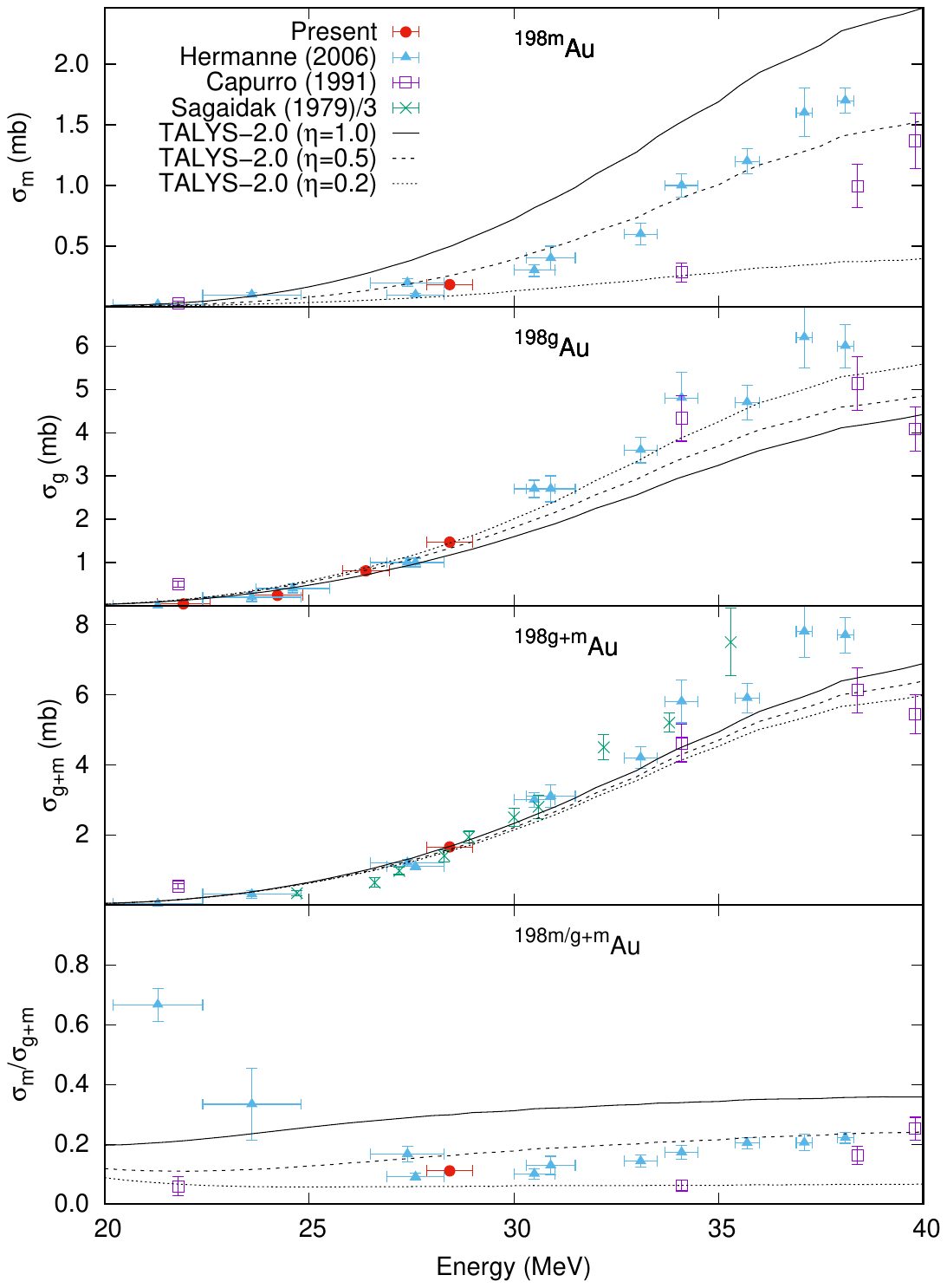}
\end{center}
\caption{
\natu Pt($\alpha$,x)$^{198m,g}$Au reaction cross sections along with their sum $\sigma_g+\sigma_m$ and ratio $\sigma_m/(\sigma_g+\sigma_m)$.
The cross sections of Sagaidak et al. are multiplied by 1/3.
The $^{198g+m}$Au production cross sections and isomeric ratios of Hermanne et al. and Capurro et al. were derived from  the isomer production cross sections originally published by them.
}
\label{fig:198Au}
\end{figure}

Figure~\ref{fig:198Au} compares the production cross sections and isomeric ratios of $^{198m,g}$Au measured in this work with those measured by Hermanne et al.~\cite{Hermanne2006}, Capurro et al.~\cite{Capurro1991} and Sagaidak et al.~\cite{Sagaidak1979}.
The data measured in the present work are also listed in Table~\ref{tab:198Au}.
We were able to determine the production cross section of $^{198m}$Au and isomeric ratio only for the first Pt foil at the highest energy.
The factor $p\lambda_m/(\lambda_m-\lambda_g)$ in Eq.~(\ref{eqn:cumgm2}) is 6.38 for $^{198}$Au, which is very high and indicating that Eq.~(\ref{eqn:cumgm2}) or (\ref{eqn:cumgm3}) cannot be used.
Use of Eq.~(\ref{eqn:cumgm3}) causes overestimation of the ground state production cross section by more than 60\% at around 30~MeV according to Table~\ref{tab:198Au}.
It is therefore obvious that use of Eq.~(\ref{eqn:cumgm1}) is obligatory for proper determination of the ground state production cross section if one determines the cross section with the activation cross section formula.

The cross sections and isomeric ratio measured by Hermanne et al. are consistent with our results, and we would conclude that Hermanne et al. consider the correction factor 6.38 in determination of their ground state production cross section.
The metastable state production cross sections reported by Capurro et al. are systematically lower than those reported by Hermanne et al.
The 215 keV $\gamma$ emission probability adopted by Capurro et al. (76.9\%) agrees with the one adopted by us,
and the reason of the deviation in their isomeric ratio is not clear.
Sagaidak et al. report only the $^{198g+m}$Au total cross sections, of which energy dependence is consistent with that of Hermanne et al. though the absolute cross sections reported by Sagaidak et al. are systematically higher by a factor of $\sim$3.
The emission probability of the 412~keV $\gamma$ line adopted by Sagaidak et al. (95.53\%) agrees with the one adopted by Hermanne et al. and us.

TALYS-2.0 gives better prediction of the isomeric ratio if we reduce the spin cutoff parameter from its default value, and $\eta\sim$0.5 looks appropriate to describe the energy dependent isomeric ratio reported by Hermanne et al.

\begin{table}[hbtp]
\caption{
$^{198}$Au production cross section (mb) and isomeric ratio $IR=\sigma_m/(\sigma_g+\sigma_m)$ as a function of the incident energy (MeV). 
}
\label{tab:198Au}
\begin{center}
\begin{tabular}{
D{.}{.}{1}
D{.}{.}{1}
D{.}{.}{3}
D{.}{.}{3}
D{.}{.}{2}
D{.}{.}{2}
D{.}{.}{2}
D{.}{.}{2}
D{.}{.}{3}
D{.}{.}{3}
}
\hline
\hline
\multicolumn{1}{c}{$E$}                 &
\multicolumn{1}{c}{$\Delta E$}          &
\multicolumn{1}{c}{$\sigma_m$}          &
\multicolumn{1}{c}{$\Delta\sigma_m$}    &
\multicolumn{1}{c}{$\sigma_g$}          &
\multicolumn{1}{c}{$\Delta\sigma_g$}    &
\multicolumn{1}{c}{$\sigma_{g+m}$}      &
\multicolumn{1}{c}{$\Delta\sigma_{g+m}$}&
\multicolumn{1}{c}{$IR$}                &
\multicolumn{1}{c}{$\Delta IR$}         \\
\hline
28.4   &0.6       &0.1823    &0.0079          &1.471     &0.016           &1.653         &0.018               &0.1103&0.0044     \\
26.4   &0.6       &          &                &0.808     &0.074           &              &                    &      &           \\
24.2   &0.6       &          &                &0.242     &0.048           &              &                    &      &           \\
21.9   &0.6       &          &                &0.045     &0.015           &              &                    &      &           \\
\hline
\hline
\end{tabular}
\end{center}
\end{table}

\subsubsection{$^{197}$Hg}
\begin{figure}[hbtp]
\begin{center}
\includegraphics[width=0.9\textwidth]{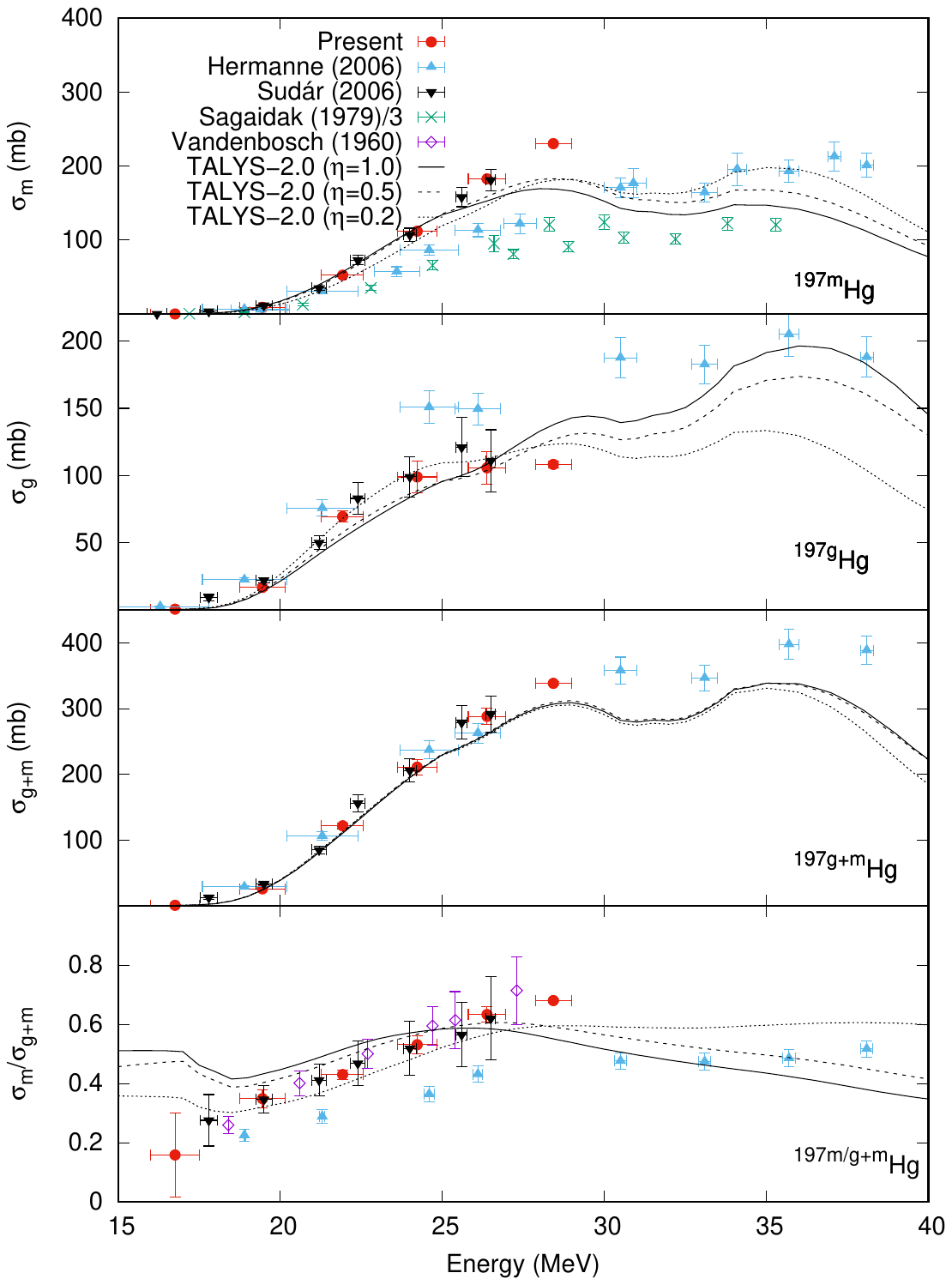}
\end{center}
\caption{
\natu Pt($\alpha$,x)$^{197m,g}$Hg reaction cross sections along with their sum $\sigma_g+\sigma_m$ and ratio $\sigma_m/(\sigma_g+\sigma_m)$.
The cross sections of Sagaidak et al. are multiplied by 1/3.
The $^{197g+m}$Hg production cross sections of Hermanne et al. and Sud\'{a}r et al. as well as the isomeric ratios of Hermanne et al. were derived from the isomer production cross sections originally published by them. 
}
\label{fig:197Hg}
\end{figure}

Figure~\ref{fig:197Hg} compares the production cross sections and isomeric ratios of $^{197m,g}$Hg measured in this work with those measured by Hermanne et al.~\cite{Hermanne2006}, Sud\'{a}r et al.~\cite{Sudar2006}, Sagaidak et al.~\cite{Sagaidak1979} and Vandenbosch et al.~\cite{Vandenbosch1960}.
The data measured in the present work are also listed in Table~\ref{tab:197Hg}.
The factor $p\lambda_m/(\lambda_m-\lambda_g)$ in Eq.~(\ref{eqn:cumgm2}) is 1.45 for $^{197}$Hg,
which is not as high as the factor for $^{198}$Au.
However,
the metastable state is more populated than the ground state at $\sim$30~MeV ($\sigma_m/\sigma_g\sim$2),
and this makes the approximation by Eq.~(\ref{eqn:cumgm3}) worse.
The ground state and metastable state production cross sections determined by us predict that use of Eq.~(\ref{eqn:cumgm3}) causes overestimation of the ground state production cross section by more than a factor of 2 at $\sim$30 MeV.
Therefore,
use of Eq.~(\ref{eqn:cumgm3}) must be avoided for proper determination of the $^{197g}$Hg production cross section.

Figure~\ref{fig:197Hg} shows an excellent agreement between our data and data published by Sud\'{a}r et al. for all four quantities.
They adopt the 134~keV $\gamma$ line rather than the 279~keV $\gamma$ line for quantification of the metastable state production.
As expected from Fig.~\ref{fig:tcdeprev}, the impact of the IT branching ratio correction on the metastable state production cross section and isomeric ratio is minor if the 134~keV $\gamma$ line is adopted, and this explains the fact that the data published by Sud\'{a}r et al. are not affected by the absence of the IT branching ratio correction.
Sud\'{a}r et al. show that they use the proper data reduction equation without an approximation, namely Eq.~(\ref{eqn:cumgm1}).
Vandenbosch et al. report only the isomeric ratios, which are again in an excellent agreement with those determined by us.
They also adopt the 134~keV $\gamma$ line for quantification of the metastable state.
The measurements by Sud\'{a}r et al. and Vandenbosch et al. aim to study the spin cutoff parameter,
and one can expect that the decay scheme is carefully considered in their determination of the isomeric ratios.
Hermanne et al. adopt not only the 134~keV $\gamma$ line but also the 279~keV $\gamma$ line for quantification of the metastable state production,
and this explains the deviation of their $^{197m}$Hg production cross section.
Sagaidak et al. report only the metastable state production cross section,
which is about three times higher than the cross sections from the other measurements as seen in the case of $^{198}$Au in Fig.~\ref{fig:198Au}.

Again TALYS-2.0 gives better prediction if we reduce the spin cutoff parameter from its default value.
Sud\'{a}r et al. performed model calculations with another reaction model code STAPRE~\cite{Uhl1976} with $\eta=$0.15, 0.20 and 0.25,
and got best fit to the isomeric ratios measured by them with $\eta=0.20$,
which also looks a good choice in our calculation for description of the isomeric ratio.

\begin{table}[hbtp]
\caption{
$^{197}$Hg production cross section (mb) and isomeric ratio $IR=\sigma_m/(\sigma_g+\sigma_m)$ as a function of the incident energy (MeV). 
An asterisk indicates presence of a value with uncertainty higher than 50\%.
}
\label{tab:197Hg}
\begin{center}
\begin{tabular}{
D{.}{.}{1}
D{.}{.}{1}
D{.}{.}{3}
D{.}{.}{3}
D{.}{.}{0}
D{.}{.}{0}
D{.}{.}{0}
D{.}{.}{0}
D{.}{.}{3}
D{.}{.}{3}
}
\hline
\hline
\multicolumn{1}{c}{$E$}                 &
\multicolumn{1}{c}{$\Delta E$}          &
\multicolumn{1}{c}{$\sigma_m$}          &
\multicolumn{1}{c}{$\Delta\sigma_m$}    &
\multicolumn{1}{c}{$\sigma_g$}          &
\multicolumn{1}{c}{$\Delta\sigma_g$}    &
\multicolumn{1}{c}{$\sigma_{g+m}$}      &
\multicolumn{1}{c}{$\Delta\sigma_{g+m}$}&
\multicolumn{1}{c}{$IR$}                &
\multicolumn{1}{c}{$\Delta IR$}         \\
\hline
28.4   &0.6       &230.41    &0.16            &108.2     &2.3             &338.6         &2.3                 &0.6805&0.0046     \\
26.4   &0.6       &182.67    &0.64            &106       &12              &288           &12                  &0.634 &0.027      \\
24.2   &0.6       &112.00    &0.31            &99        &12              &211           &12                  &0.531 &0.030      \\
21.9   &0.6       &52.39     &0.28            &69.4      &4.3             &121.8         &4.3                 &0.430 &0.015      \\
19.5   &0.7       &8.978     &0.088           &16.8      &2.3             &25.7          &2.3                 &0.349 &0.031      \\
16.8   &0.8       &0.1122    &0.0064          &*         &*               &*             &*                   &*     &*          \\
10.2   &1.0       &          &                &*         &*               &              &                    &      &           \\
\hline
\hline
\end{tabular}
\end{center}
\end{table}

\subsubsection{$^{195}$Hg}
\begin{figure}[hbtp]
\begin{center}
\includegraphics[width=0.9\textwidth]{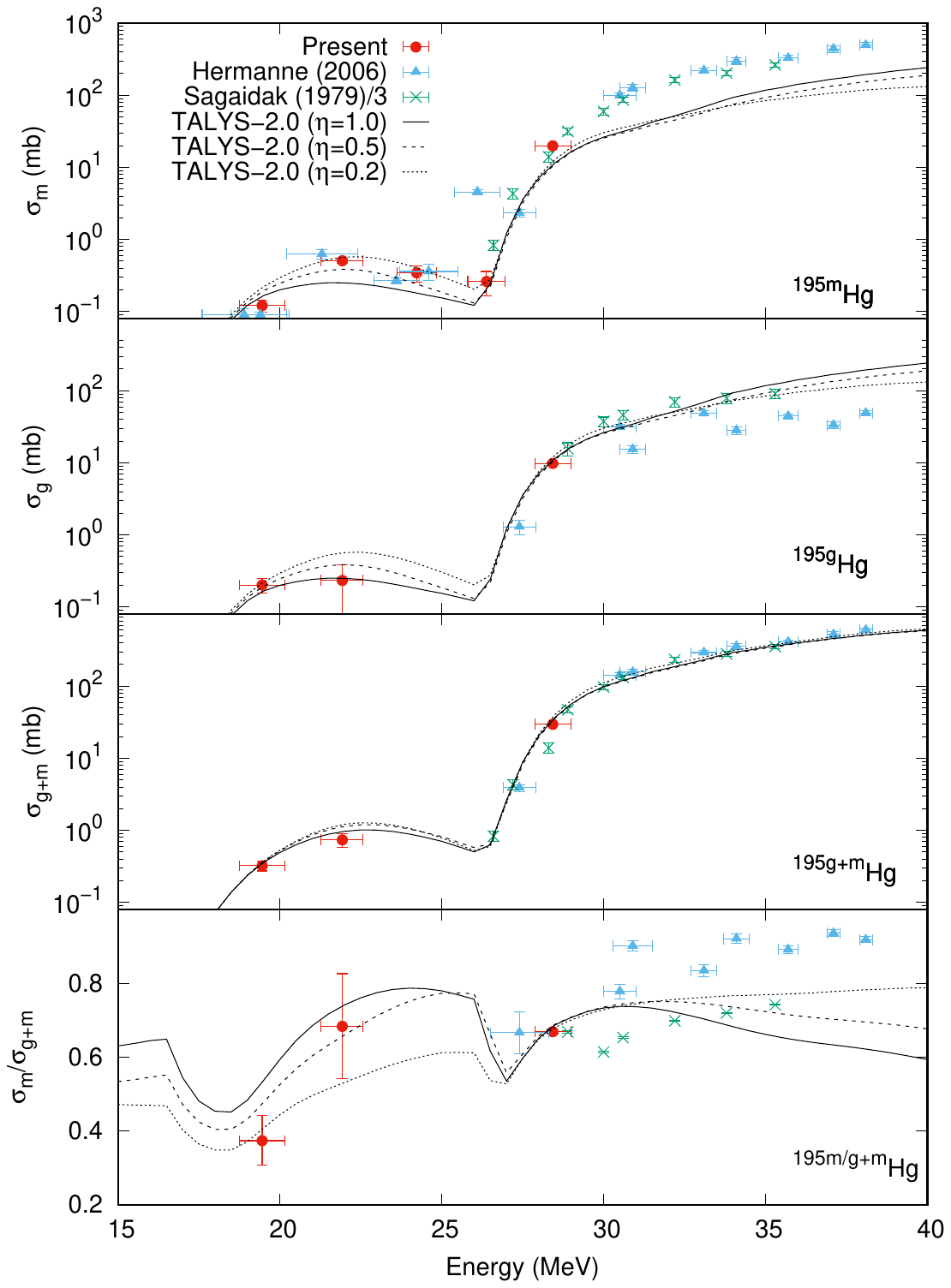}
\end{center}
\caption{
\natu Pt($\alpha$,x)$^{195m,g}$Hg reaction cross sections along with their sum $\sigma_g+\sigma_m$ and ratio $\sigma_m/(\sigma_g+\sigma_m)$.
The cross sections of Sagaidak et al. are multiplied by 1/3.
The $^{195g+m}$Hg production cross sections and isomeric ratios of Hermanne et al. and Sagaidak et al. were derived from the isomer production cross sections originally published by them.
}
\label{fig:195Hg}
\end{figure}

Figure~\ref{fig:195Hg} compares the production cross sections and isomeric ratios of $^{195m,g}$Hg measured in this work with those measured by Hermanne et al.~\cite{Hermanne2006} and Sagaidak et al.~\cite{Sagaidak1979}.
The data measured in the present work are also listed in Table~\ref{tab:195Hg}.
At 28.4~MeV, the metastable state production cross section multiplied by the corrected IT and EC/$\beta^+$ branching ratios are 9.7 and 10.2~mb, respectively, and they are also close to the ground state production cross section (9.8~mb).
This explains the agreement of their partial decay curves of the 98~keV $\gamma$ line following the $^{195}$Au half-life for cooling longer than 100~h in Fig.~\ref{fig:decaycurve}.
The half-life of the metastable state is longer than the half-life of the ground state.
Hermanne et al. measured the ground state activity shortly after the end of bombardment to reduce the correction due to isomeric transition.
Therefore, one would expect our revision in the IT branching ratio does not have a major impact on the ground state production cross section measured by them.
They determined the metastable state production cross section by measuring the 262, 388 and 560~keV $\gamma$ lines, which all follow EC/$\beta^+$ decay of $^{195m}$Hg,
and are subject to $\sim$10\% increase of the $\gamma$ emission probabilities introduced in Table~\ref{tab:decaydatarev}.
Reduction of the metastable state production cross sections measured by Hermanne et al. due to revision of the IT branching ratio is very small in logarithmic scale and not visible in Fig.~\ref{fig:195Hg} even if their metastable state production cross sections renormalized with the corrected IT branching ratio are plotted together.
The energy dependence of the metastable state and ground state production cross sections measured by Sagaidak et al. is close to the other measurements though their cross sections are again about a factor of $\sim$3 systematically higher.

Reduction of the spin cutoff parameter improves TALYS-2.0 results in better prediction for the isomeric ratio though we cannot choose one from $\eta=0.5$ and 0.2 for better description of the energy dependence observed by us.

\begin{table}[hbtp]
\caption{
$^{195}$Hg production cross section (mb) and isomeric ratio $IR=\sigma_m/(\sigma_g+\sigma_m)$ as a function of the incident energy (MeV). 
An asterisk indicates presence of a value with uncertainty higher than 50\%.
}
\label{tab:195Hg}
\begin{center}
\begin{tabular}{
D{.}{.}{1}
D{.}{.}{1}
D{.}{.}{2}
D{.}{.}{2}
D{.}{.}{2}
D{.}{.}{2}
D{.}{.}{2}
D{.}{.}{2}
D{.}{.}{3}
D{.}{.}{3}
}
\hline
\hline
\multicolumn{1}{c}{$E$}                 &
\multicolumn{1}{c}{$\Delta E$}          &
\multicolumn{1}{c}{$\sigma_m$}          &
\multicolumn{1}{c}{$\Delta\sigma_m$}    &
\multicolumn{1}{c}{$\sigma_g$}          &
\multicolumn{1}{c}{$\Delta\sigma_g$}    &
\multicolumn{1}{c}{$\sigma_{g+m}$}      &
\multicolumn{1}{c}{$\Delta\sigma_{g+m}$}&
\multicolumn{1}{c}{$IR$}                &
\multicolumn{1}{c}{$\Delta IR$}         \\
\hline
28.4   &0.6       &19.92     &0.13            &9.83      &0.23            &29.75         &0.27                &0.6694&0.0055     \\
26.4   &0.6       &0.261     &0.098           &          &                &              &                    &      &           \\
24.2   &0.6       &0.344     &0.085           &          &                &              &                    &      &           \\
21.9   &0.6       &0.507     &0.042           &*         &*               &0.74          &0.16                &0.68  &0.14       \\
19.5   &0.7       &0.121     &0.022           &0.202     &0.044           &0.323         &0.050               &0.374 &0.067      \\
\hline
\hline
\end{tabular}
\end{center}
\end{table}

\subsubsection{Other product nuclides}
\begin{figure}[hbtp]
\begin{center}
\includegraphics[width=0.8\textwidth]{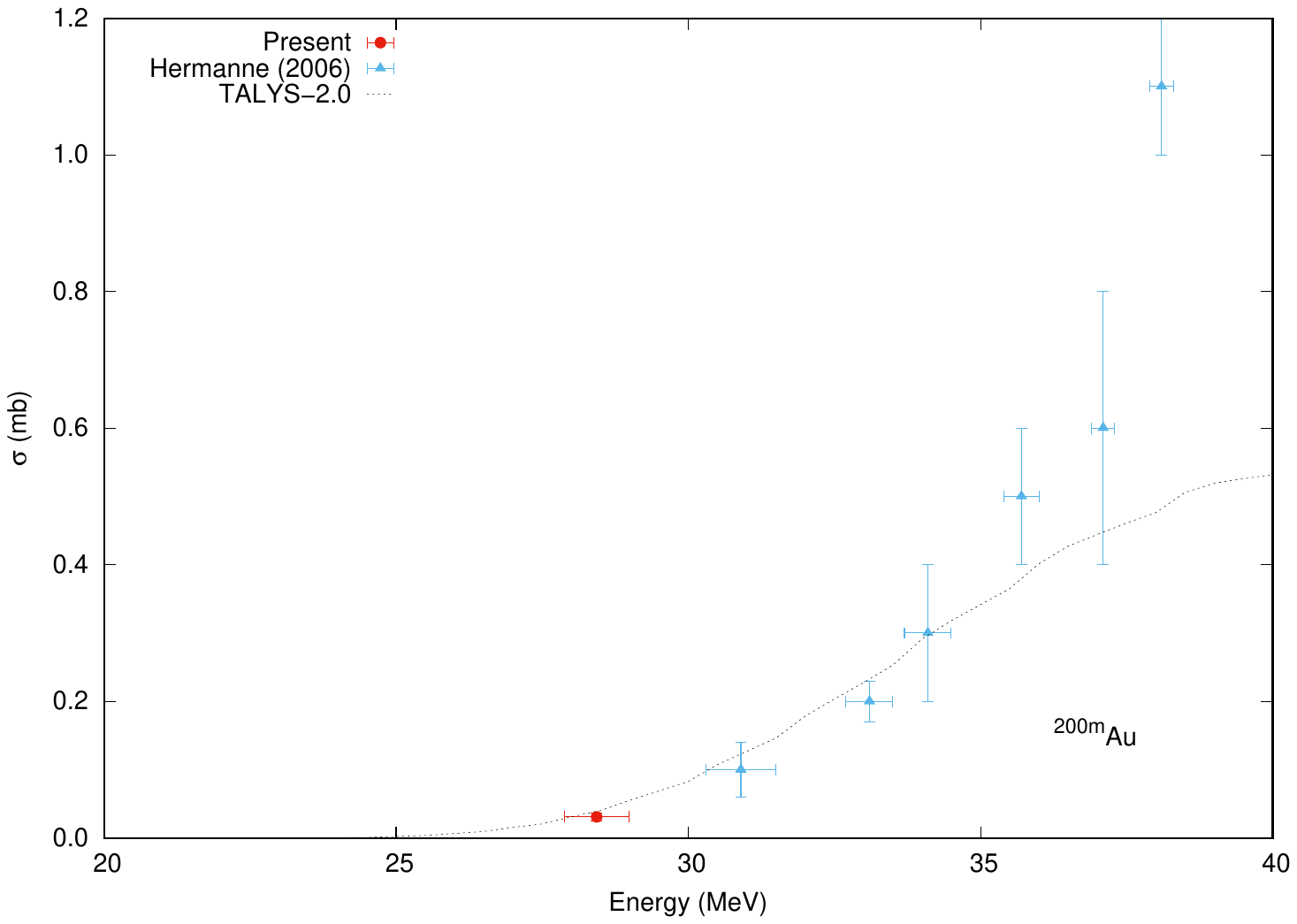}
\end{center}
\caption{
\natu Pt($\alpha$,x)$^{200m}$Au reaction cross sections.
}
\label{fig:200mAu}
\end{figure}

\begin{figure}[hbtp]
\begin{center}
\includegraphics[width=0.8\textwidth]{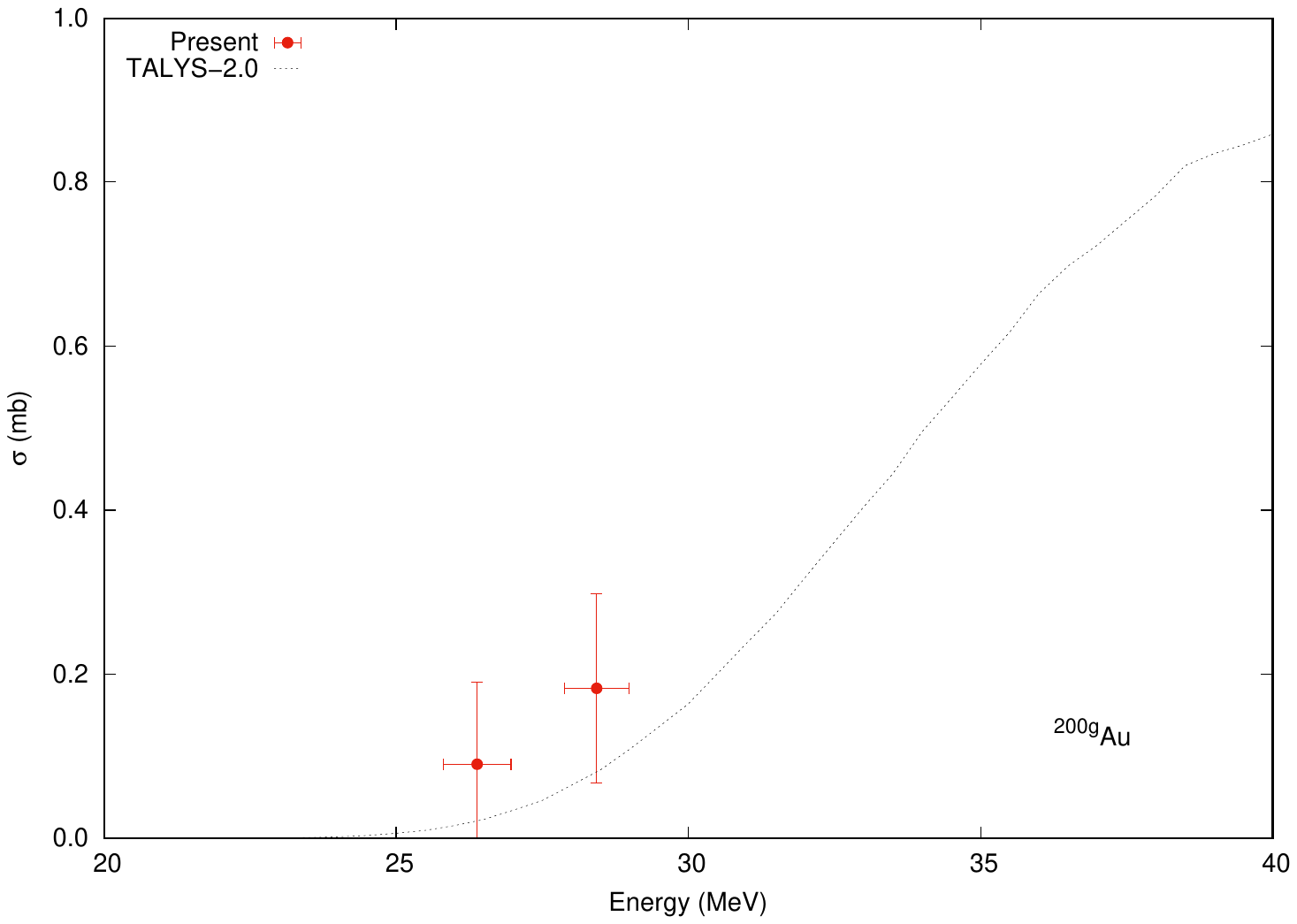}
\end{center}
\caption{
\natu Pt($\alpha$,x)$^{200g}$Au reaction cross sections.
}
\label{fig:200gAu}
\end{figure}

\begin{figure}[hbtp]
\begin{center}
\includegraphics[width=0.8\textwidth]{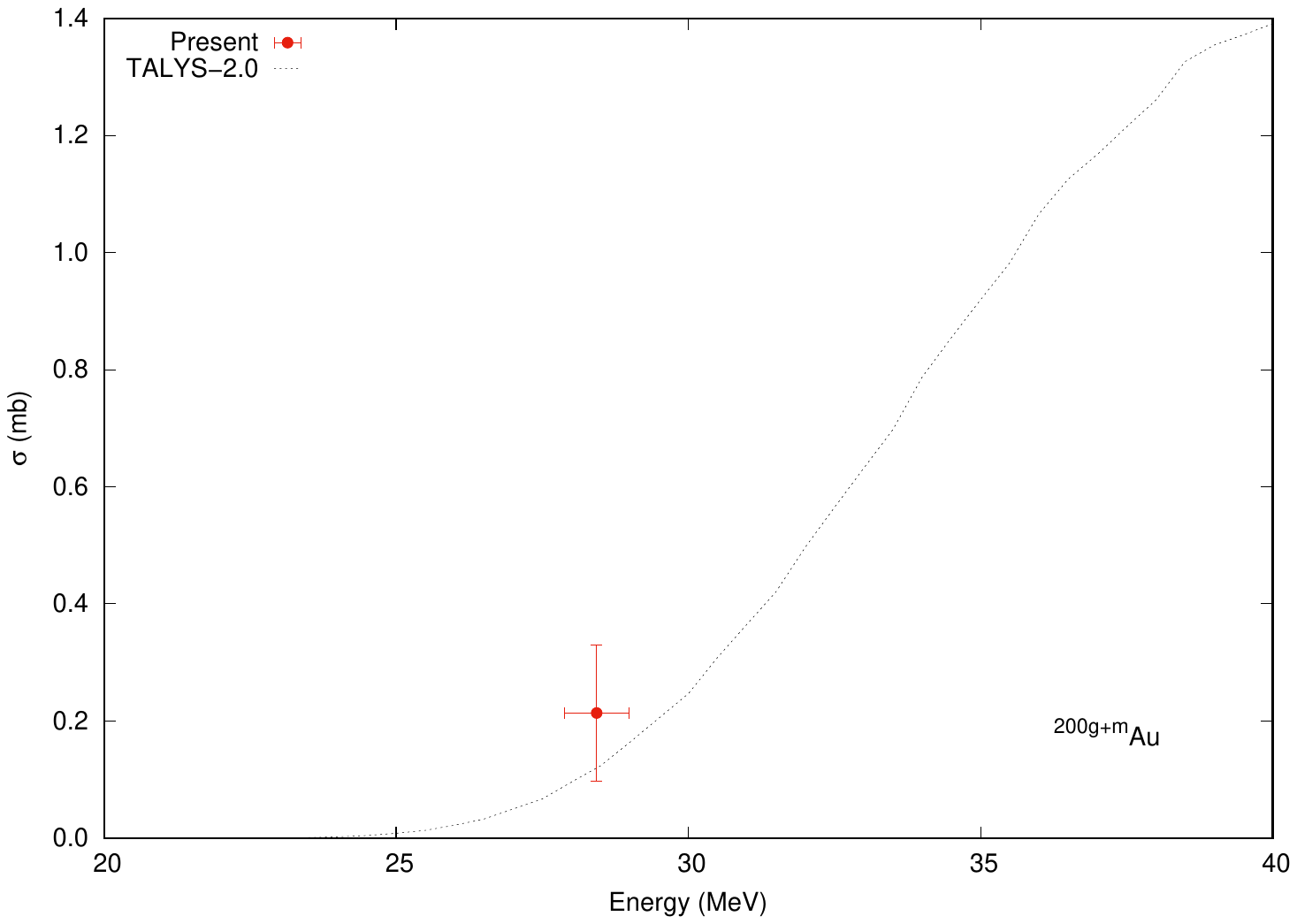}
\end{center}
\caption{
\natu Pt($\alpha$,x)$^{200g+m}$Au reaction cross sections.
}
\label{fig:200Au}
\end{figure}

\begin{figure}[hbtp]
\begin{center}
\includegraphics[width=0.8\textwidth]{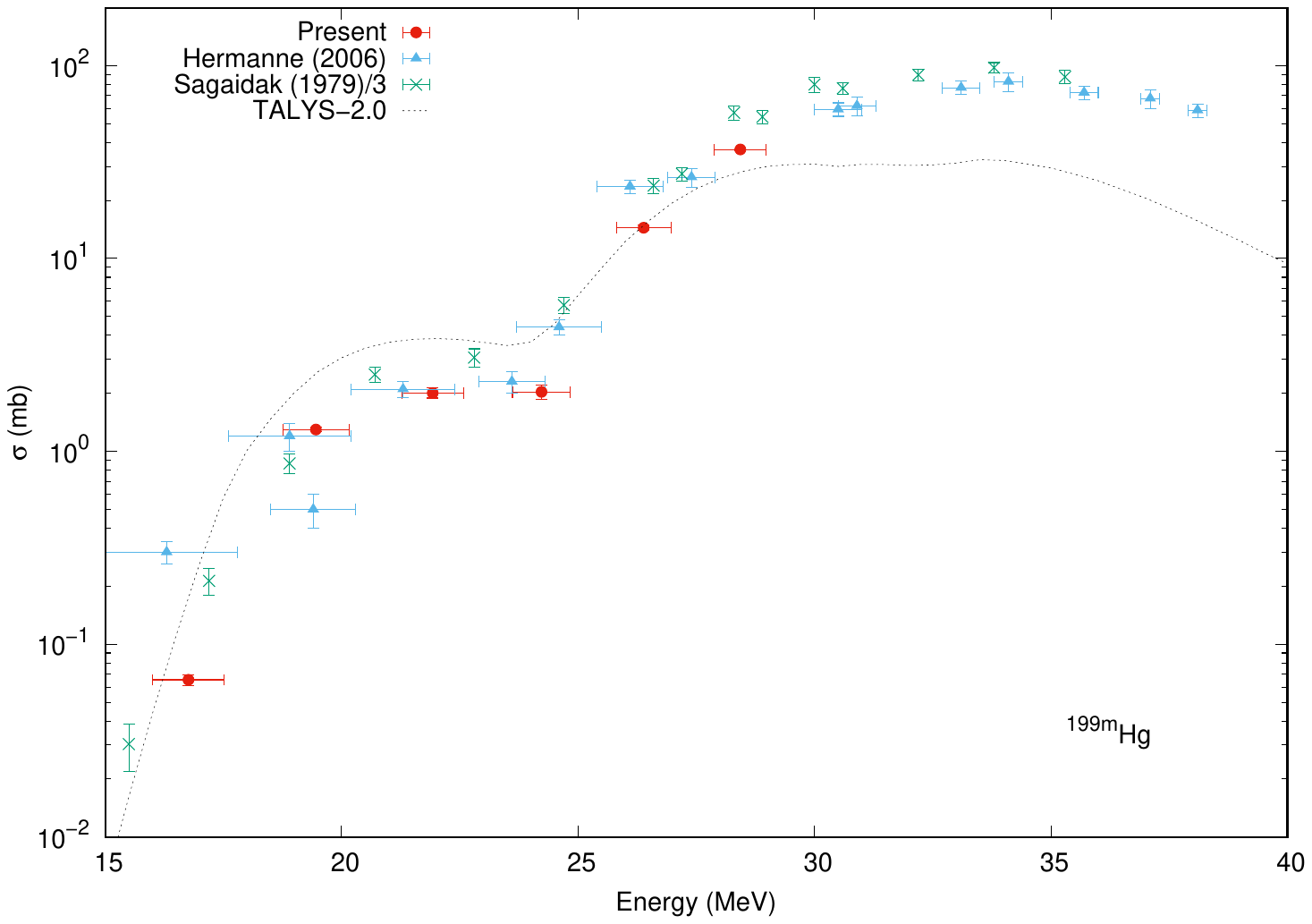}
\end{center}
\caption{
\natu Pt($\alpha$,x)$^{199m}$Hg reaction cross sections.
}
\label{fig:199mHg}
\end{figure}

\begin{figure}[hbtp]
\begin{center}
\includegraphics[width=0.8\textwidth]{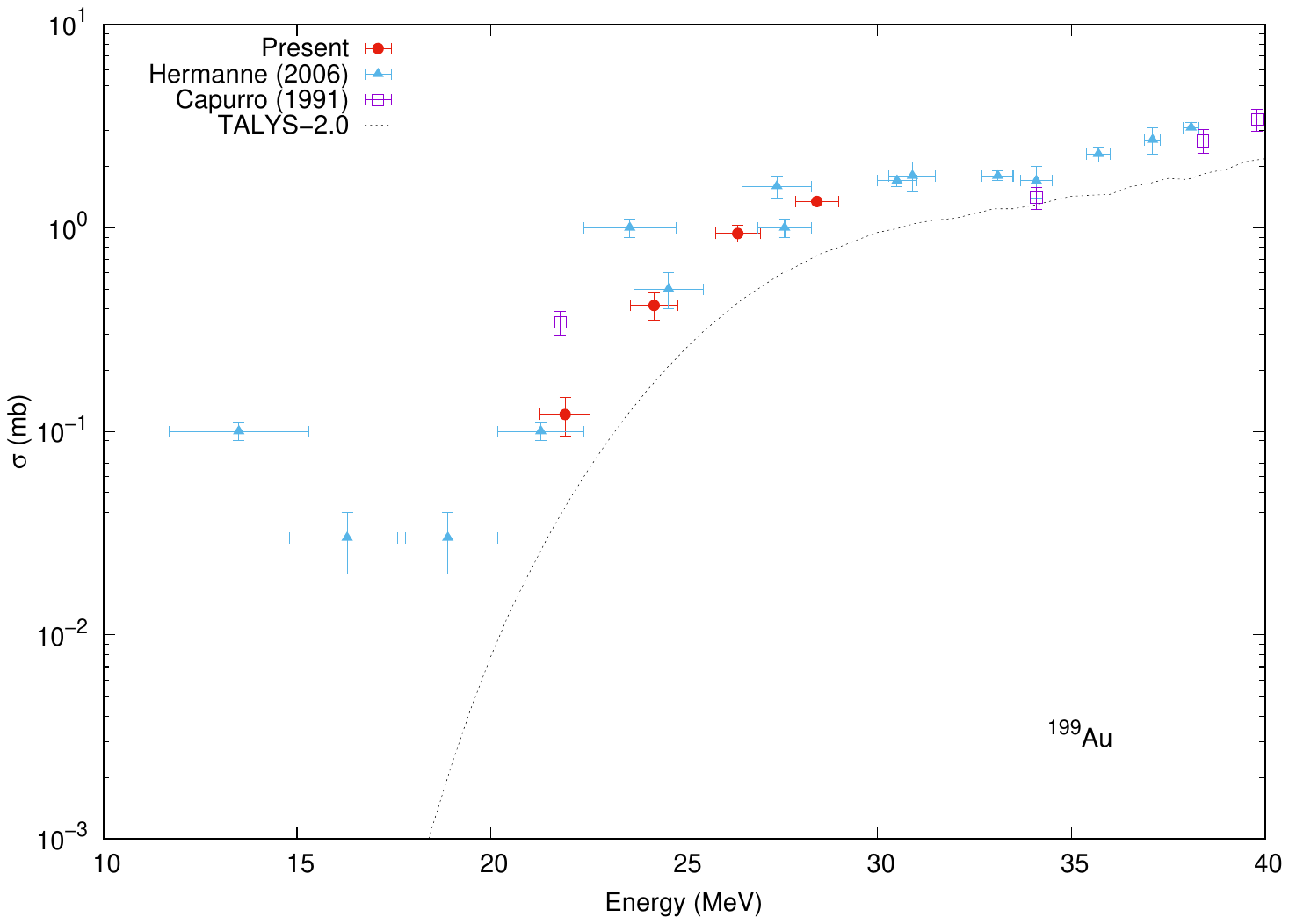}
\end{center}
\caption{
\natu Pt($\alpha$,x)$^{199}$Au reaction cross sections.
}
\label{fig:199Au}
\end{figure}

\begin{figure}[hbtp]
\begin{center}
\includegraphics[width=0.8\textwidth]{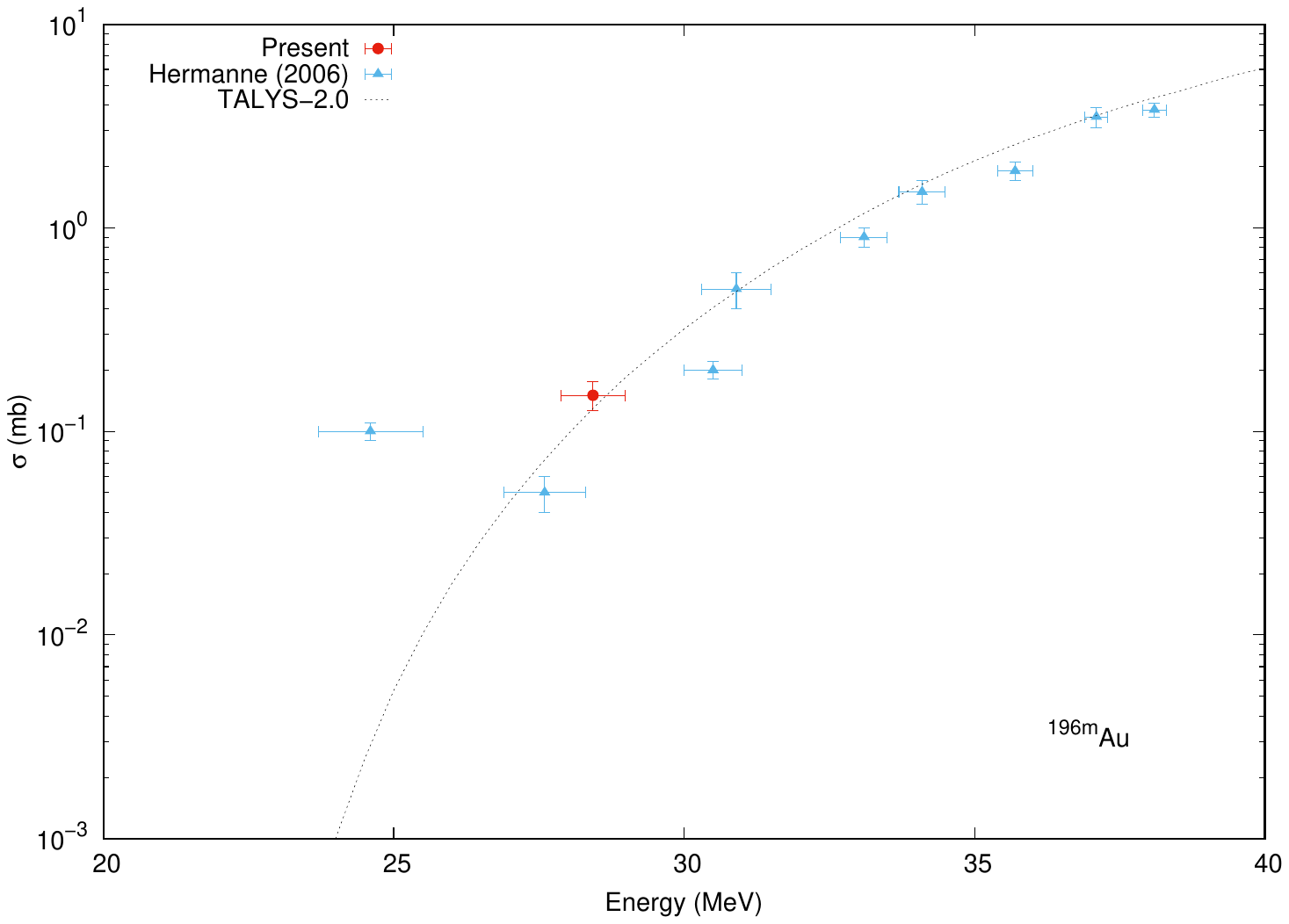}
\end{center}
\caption{
\natu Pt($\alpha$,x)$^{196m}$Au reaction cross sections.
}
\label{fig:196mAu}
\end{figure}

\begin{figure}[hbtp]
\begin{center}
\includegraphics[width=0.8\textwidth]{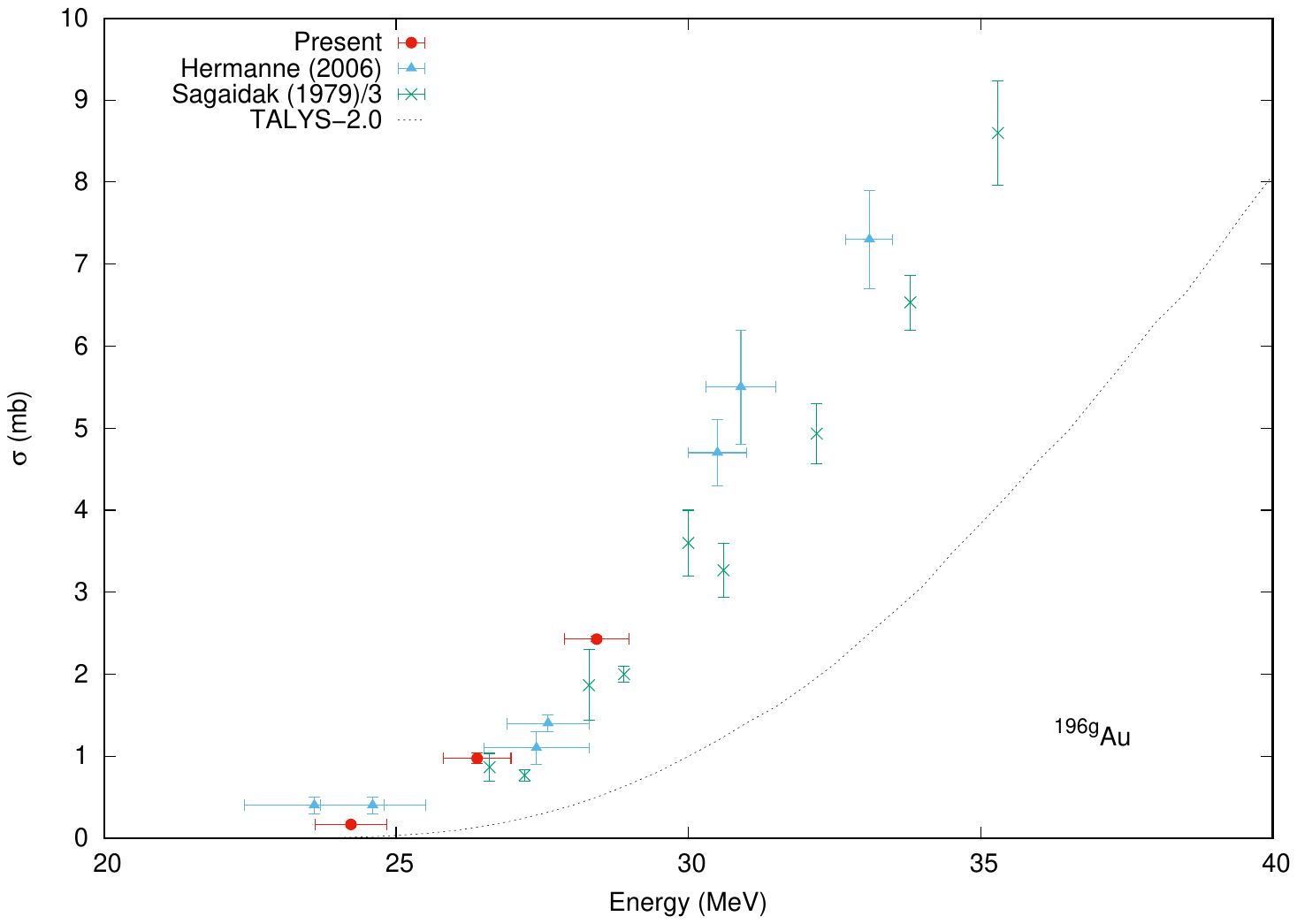}
\end{center}
\caption{
\natu Pt($\alpha$,x)$^{196g}$Au reaction cross sections.
}
\label{fig:196gAu}
\end{figure}

\begin{figure}[hbtp]
\begin{center}
\includegraphics[width=0.8\textwidth]{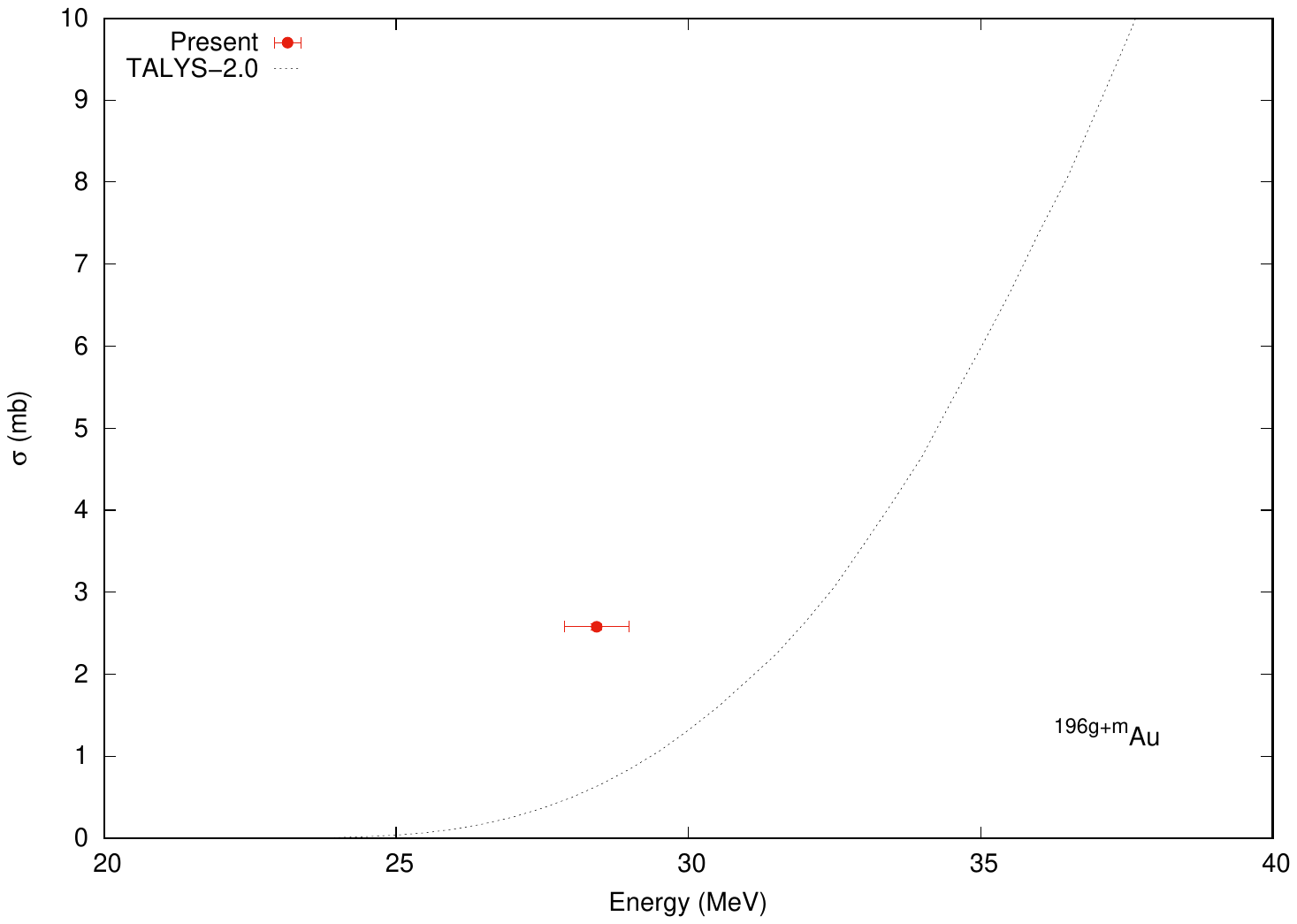}
\end{center}
\caption{
\natu Pt($\alpha$,x)$^{196g+m}$Au reaction cross sections.
}
\label{fig:196Au}
\end{figure}

\begin{figure}[hbtp]
\begin{center}
\includegraphics[width=0.8\textwidth]{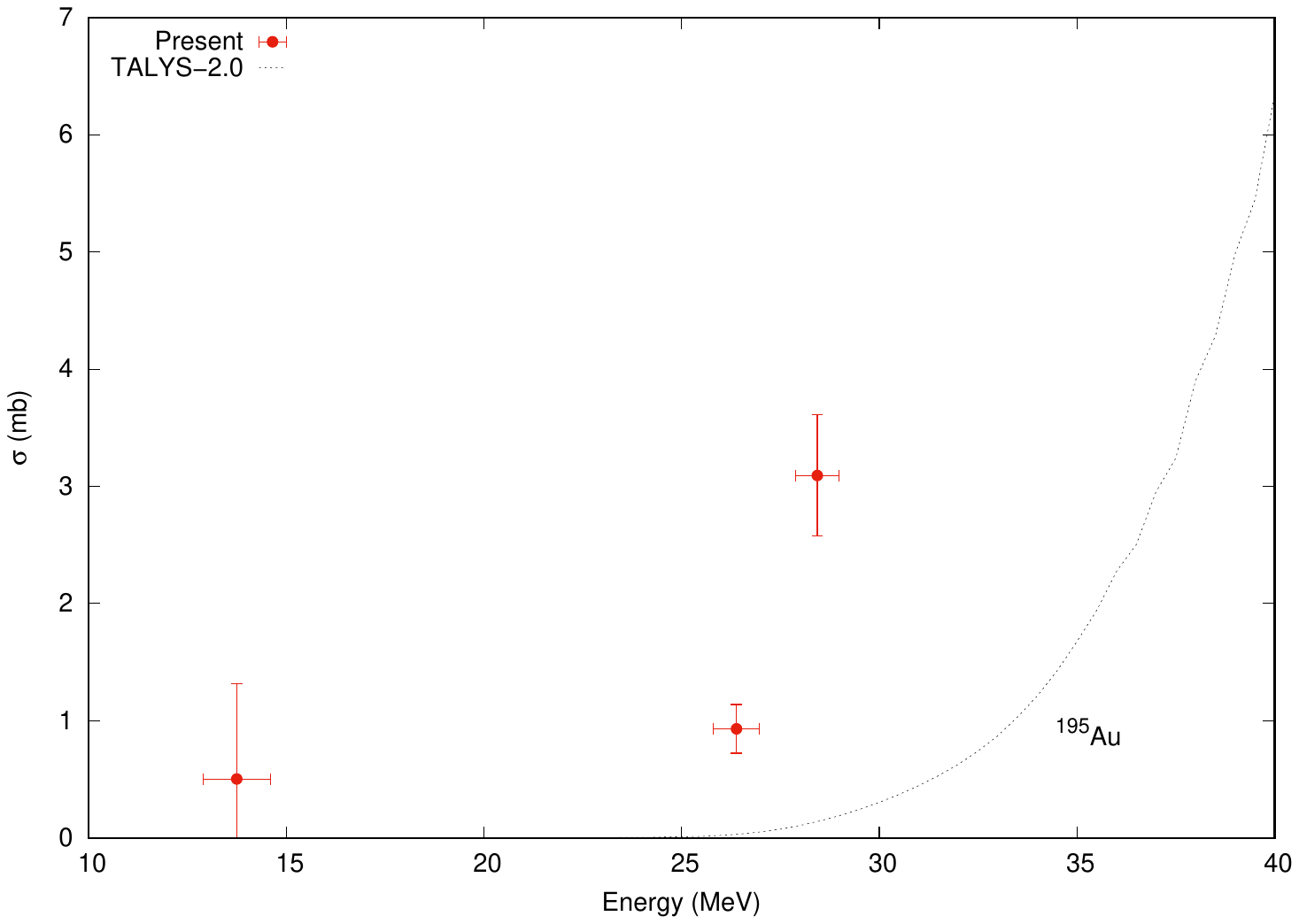}
\end{center}
\caption{
\natu Pt($\alpha$,x)$^{195}$Au reaction cross sections.
}
\label{fig:195Au}
\end{figure}

\begin{figure}[hbtp]
\begin{center}
\includegraphics[width=0.8\textwidth]{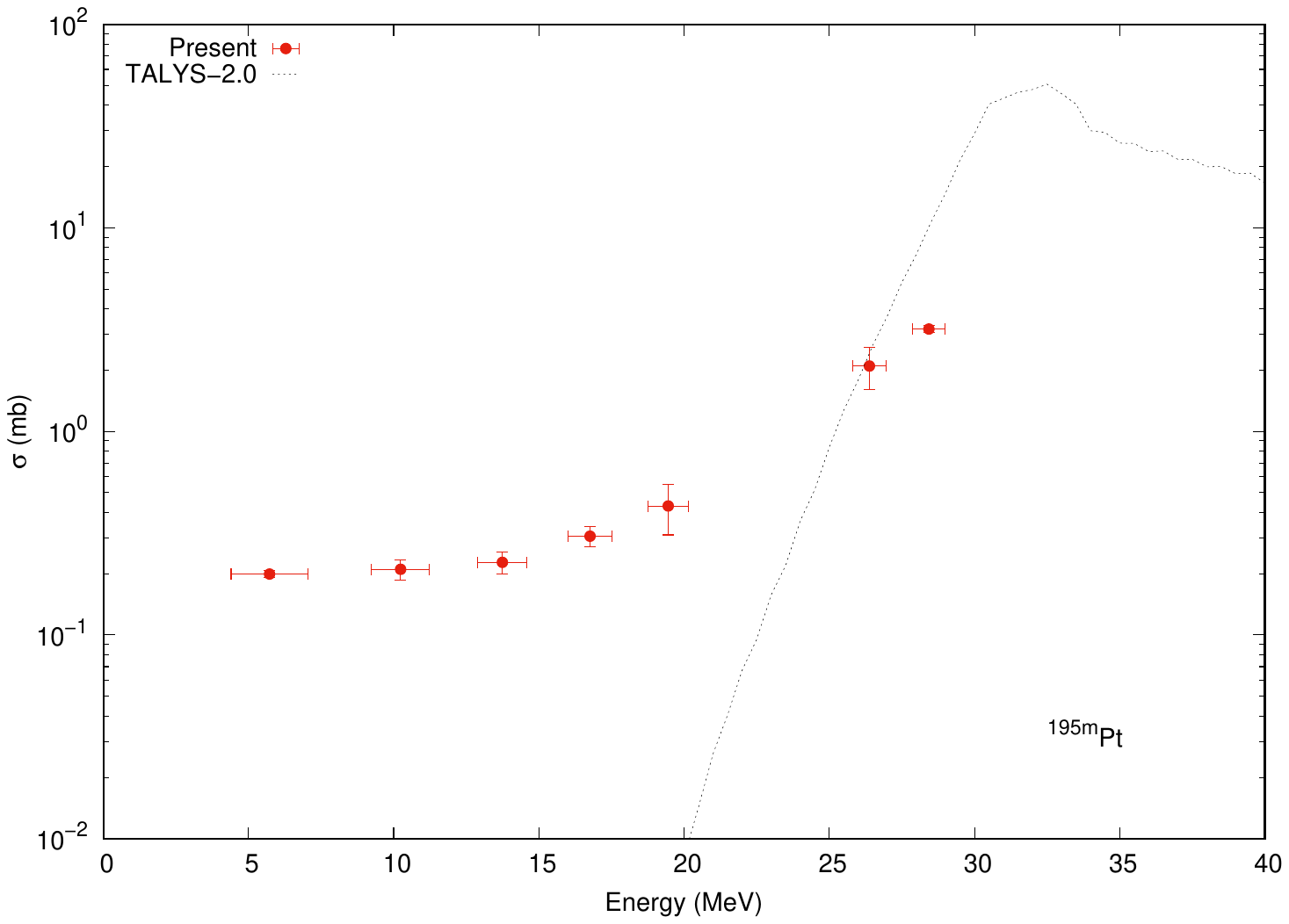}
\end{center}
\caption{
\natu Pt($\alpha$,x)$^{195m}$Pt reaction cross sections.
The present result below 20~MeV must be seen with caution as discussed in the main text.
}
\label{fig:195mPt}
\end{figure}

\begin{figure}[hbtp]
\begin{center}
\includegraphics[width=0.8\textwidth]{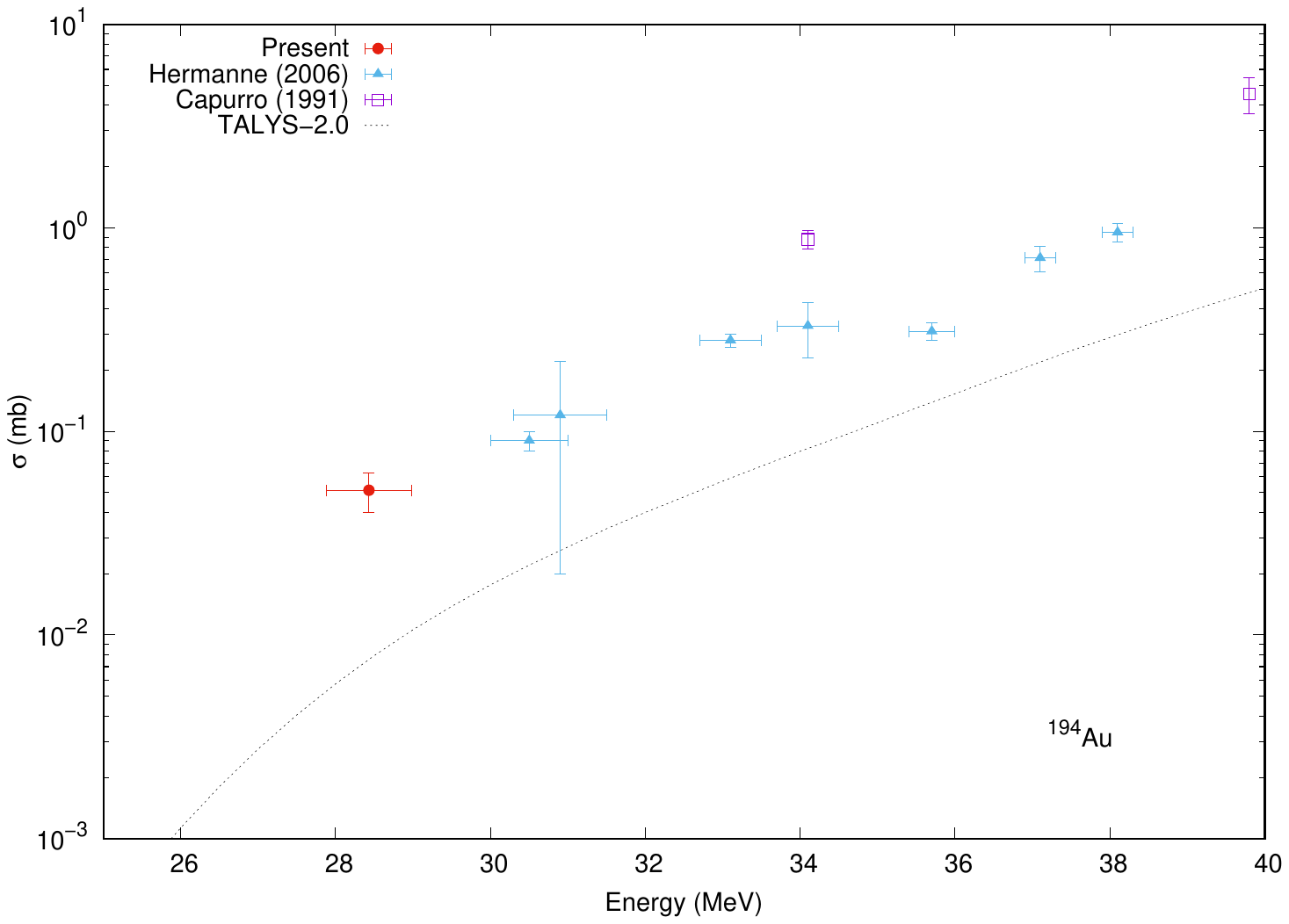}
\end{center}
\caption{
\natu Pt($\alpha$,x)$^{194}$Au reaction cross sections.
}
\label{fig:194Au}
\end{figure}

\begin{figure}[hbtp]
\begin{center}
\includegraphics[width=0.8\textwidth]{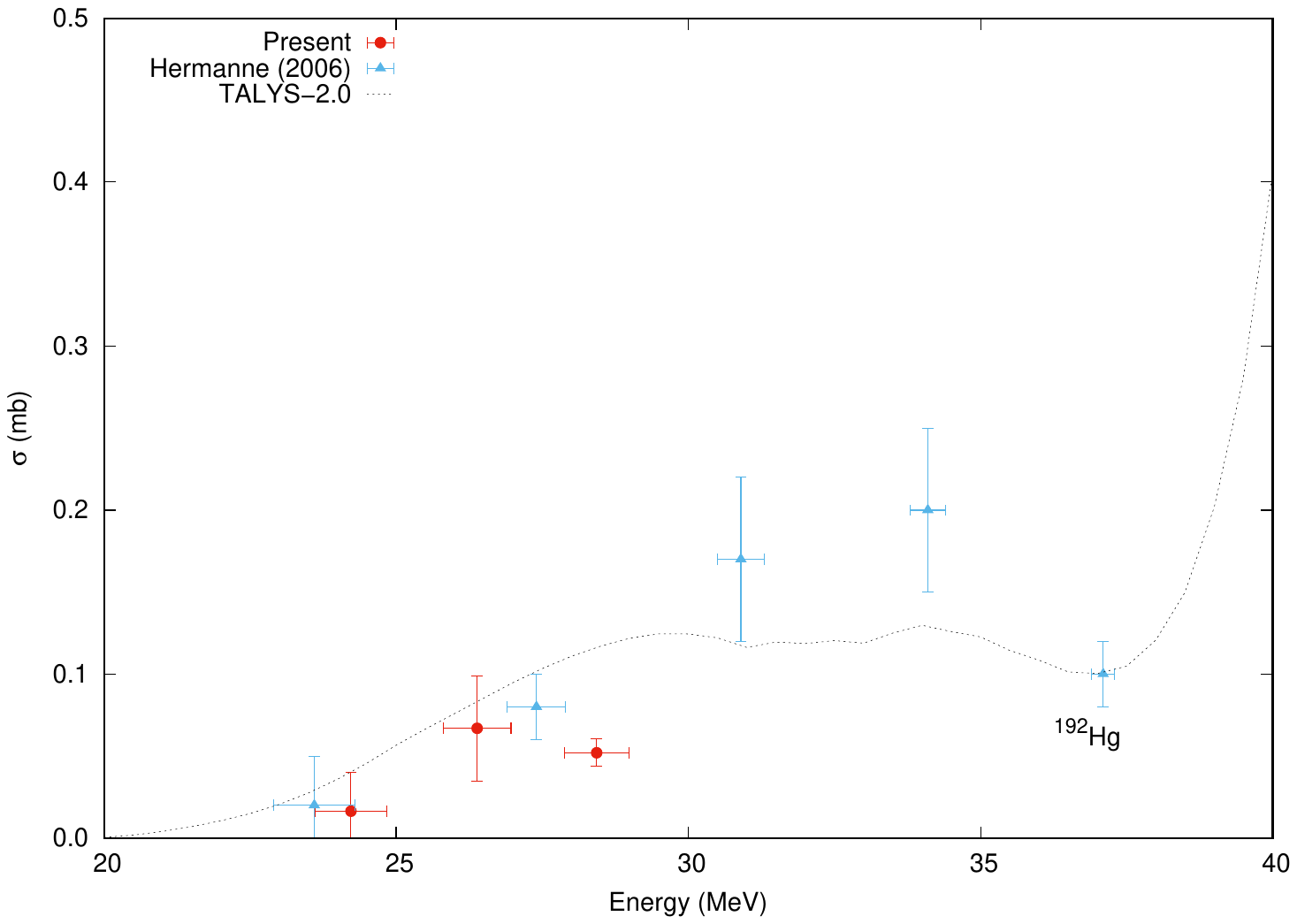}
\end{center}
\caption{
\natu Pt($\alpha$,x)$^{192}$Hg reaction cross sections.
}
\label{fig:192Hg}
\end{figure}

The production cross sections of $^{200m,g,g+m}$Au, $^{199m}$Hg, $^{199}$Au, $^{196m,g,g+m}$Au, $^{195}$Au, $^{195m}$Pt, $^{194}$Au and $^{192}$Hg are plotted in Figs.~\ref{fig:200mAu} to \ref{fig:192Hg} and listed in Tables~\ref{tab:200Au199Hg} to \ref{tab:195Au195Pt194Au192Hg}.
Similar to the $^{197}$Hg and $^{195}$Hg cases,
the cross sections reported by Sagaidak et al. are always a factor of $\sim$3 systematically higher than those measured by the present and other experiments.
The newly measured cross sections are consistent with the cross sections from the most recent measurement by Hermanne et al. in general.
However, many excitation functions measured in the present work do not cover the energy regions of their peaks,
and additional measurements with $\alpha$-particle energies higher than the present ones are necessary to confirm the measurement done by Hermanne et al.

The deviation between the excitation functions from the present work and TALYS-2.0 for $^{195m}$Pt below 20~MeV is remarkable.
The threshold energy of the $^{195}$Pt($\alpha$,$\alpha'$)$^{195m}$Pt reaction calculated based on the atomic masses is less than 1~MeV.
But the Coulomb barrier height between $\alpha$-particle and $^{195}$Pt is about 23~MeV assuming spherical shape with nucleon radius of 1.3~fm for both ejectile and residual.
The energy dependence predicted by TALYS-2.0 looks reasonable in this regard, and our $^{195m}$Pt production cross sections below 20~MeV must be seen with caution.

\begin{table}[hbtp]
\caption{
$^{200m,g,g+m}$Au and $^{199m}$Hg production cross sections (mb) as a function of the incident energy (MeV). 
An asterisk indicates presence of a value with uncertainty higher than 50\%.
}
\label{tab:200Au199Hg}
\begin{center}
\begin{tabular}{
D{.}{.}{1}
D{.}{.}{1}
D{.}{.}{2}
D{.}{.}{2}
D{.}{.}{2}
D{.}{.}{2}
D{.}{.}{2}
D{.}{.}{2}
c
D{.}{.}{3}
D{.}{.}{3}
}
\hline
\hline
       &          &\multicolumn{6}{c}{$^{200}$Au}                                                             &~&\multicolumn{2}{c}{$^{199}$Hg}\\
\cline{3-8}
\cline{10-11}
\multicolumn{1}{c}{$E$}                  &
\multicolumn{1}{c}{$\Delta E$}           &
\multicolumn{1}{c}{$\sigma_m$}           &
\multicolumn{1}{c}{$\Delta\sigma_m$}     &
\multicolumn{1}{c}{$\sigma_g$}           &
\multicolumn{1}{c}{$\Delta\sigma_g$}     &
\multicolumn{1}{c}{$\sigma_{g+m}$}       &
\multicolumn{1}{c}{$\Delta\sigma_{g+m}$} &
\multicolumn{1}{c}{}                     &
\multicolumn{1}{c}{$\sigma_m$}           &
\multicolumn{1}{c}{$\Delta \sigma_m$}    \\
\hline
28.4   &0.6       &0.0311    &0.0062          &*         &*               &*             &*                   & &36.76       &0.45             \\
26.4   &0.6       &          &                &*         &*               &              &                    & &14.46       &0.40             \\
24.2   &0.6       &          &                &          &                &              &                    & &2.04        &0.17             \\
21.9   &0.6       &          &                &          &                &              &                    & &2.01        &0.12             \\
19.5   &0.7       &          &                &          &                &              &                    & &1.298       &0.032            \\
16.8   &0.8       &          &                &          &                &              &                    & &0.0654      &0.0042           \\
\hline
\hline
\end{tabular}
\end{center}
\end{table}

\begin{table}[hbtp]
\caption{
$^{199}$Au and $^{196m,g,g+m}$Au production cross sections (mb) as a function of the incident energy (MeV). 
}
\label{tab:199Au196Au}
\begin{center}
\begin{tabular}{
D{.}{.}{1}
D{.}{.}{1}
D{.}{.}{2}
D{.}{.}{2}
c
D{.}{.}{2}
D{.}{.}{2}
D{.}{.}{2}
D{.}{.}{2}
D{.}{.}{2}
D{.}{.}{2}
}
\hline
\hline
       &          &\multicolumn{2}{c}{$^{199}$Au}&~&\multicolumn{6}{c}{$^{196}$Au}                                                         \\
\cline{3-4}
\cline{6-11}
\multicolumn{1}{c}{$E$}                      &
\multicolumn{1}{c}{$\Delta E$}               &
\multicolumn{1}{c}{$\sigma$}                 &
\multicolumn{1}{c}{$\Delta \sigma$}          &
\multicolumn{1}{c}{}                         &
\multicolumn{1}{c}{$\sigma_m$}               &
\multicolumn{1}{c}{$\Delta\sigma_m$}         &
\multicolumn{1}{c}{$\sigma_g$}               &
\multicolumn{1}{c}{$\Delta\sigma_g$}         &
\multicolumn{1}{c}{$\sigma_{g+m}$}           &
\multicolumn{1}{c}{$\Delta\sigma_{g+m}$}     \\
\hline                                         
28.4   &0.6       &1.346     &0.016          & &0.150     &0.025           &2.427     &0.030           &2.578         &0.039               \\
26.4   &0.6       &0.938     &0.090          & &          &                &0.973     &0.065           &              &                    \\
24.2   &0.6       &0.416     &0.063          & &          &                &0.168     &0.032           &              &                    \\
21.9   &0.6       &0.121     &0.026          & &          &                &          &                &              &                    \\
\hline
\hline
\end{tabular}
\end{center}
\end{table}

\begin{table}[hbtp]
\caption{
$^{195}$Au and $^{195m}$Pt, $^{194}$Au and $^{192}$Hg production cross sections (mb) as a function of the incident energy (MeV).
The $^{195m}$Pt production cross sections below 20~MeV must be seen with caution as discussed in the main text and therefore italicized.
An asterisk indicates presence of a value with uncertainty higher than 50\%.
}
\label{tab:195Au195Pt194Au192Hg}
\begin{center}
\begin{tabular}{
D{.}{.}{1}
D{.}{.}{1}
D{.}{.}{1}
D{.}{.}{1}
c
l
l
c
D{.}{.}{2}
D{.}{.}{2}
c
D{.}{.}{3}
D{.}{.}{3}
}
\hline
\hline
       &          &\multicolumn{2}{c}{$^{195}$Au}&~&\multicolumn{2}{c}{$^{195}$Pt}&~&\multicolumn{2}{c}{$^{194}$Au}&~&\multicolumn{2}{c}{$^{192}$Hg}\\
\cline{3-4}
\cline{6-7}
\cline{9-10}
\cline{12-13}
\multicolumn{1}{c}{$E$}             &
\multicolumn{1}{c}{$\Delta E$}      &
\multicolumn{1}{c}{$\sigma$}        &
\multicolumn{1}{c}{$\Delta \sigma$} &
\multicolumn{1}{c}{}                &
\multicolumn{1}{c}{$\sigma_m$}      &
\multicolumn{1}{c}{$\Delta\sigma_m$}&
\multicolumn{1}{c}{}                &
\multicolumn{1}{c}{$\sigma$}        &
\multicolumn{1}{c}{$\Delta\sigma$}  &
\multicolumn{1}{c}{}                &
\multicolumn{1}{c}{$\sigma$}        &
\multicolumn{1}{c}{$\Delta\sigma$}  \\
\hline
28.4   &0.6       &3.09      &0.52           & &3.18           &0.12            & &0.051   &0.011         & &0.0520  &0.0083        \\
26.4   &0.6       &0.93      &0.21           & &2.09           &0.49            & &        &              & &0.067   &0.032         \\
24.2   &0.6       &          &               & &               &                & &        &              & &*       &*             \\
19.5   &0.7       &          &               & &\textit{0.43}  &\textit{0.12}   & &        &              & &        &              \\
16.8   &0.8       &          &               & &\textit{0.306} &\textit{0.036}  & &        &              & &        &              \\
13.7   &0.9       &*         &*              & &\textit{0.227} &\textit{0.028}  & &        &              & &        &              \\
10.2   &1.0       &          &               & &\textit{0.210} &\textit{0.024}  & &        &              & &        &              \\
 5.7   &1.3       &          &               & &\textit{0.1994}&\textit{0.0086} & &        &              & &        &              \\
\hline
\hline
\end{tabular}
\end{center}
\end{table}

\clearpage

\section{Summary}
The independent production cross sections and their isomeric ratios for the radionuclides produced by irradiation of natural platinum by $\alpha$-particles up to 29~MeV were measured.
To resolve the observed cooling time dependence of the isomeric ratios of $^{197}$Hg and $^{195}$Hg determined with the conventional activation cross section formula and the decay data in the ENSDF library,
we performed simultaneous decay curve analysis at 29~MeV with the isomeric transition branching ratios of $^{197}$Hg and $^{195}$Hg as well as independent production cross sections as adjustable parameters.
The observed cooling time dependence was resolved by applying the isomeric transition branching ratios adjusted in the decay curve analysis.
We found that omission of the factor $p\lambda_m/(\lambda_m-\lambda_g)$ in determination of $\sigma_g$ from the ground state activity may introduce more than 60\% overestimation for $^{198}$Au production around 30~MeV.

We also derived the independent production cross sections of various nuclides with the newly determined $^{197,195}$Hg isomeric transition branching ratios at lower incident energies by using the simultaneous decay curve analysis.
The energy dependent $^{197}$Hg isomeric ratio measured by us shows an excellent agreement with those measured by Sud\'{a}r et al. and Vandenbosch et al.
Overall,
our cross sections are consistent with those measured by Hermanne et al.,
while the cross sections measured by Sagaidak et al. are always systematically higher by a factor of $\sim$3 than the cross sections measured by us though their energy dependence is often similar to ours.

The generalized activation cross section formula Eq.~(\ref{eqn:formula6}) shows that one can determine the independent production cross section by subtraction of the precursor decay contribution following the decay chain.
However,
the subtraction procedure adds an uncertainty,
and it is not practical to repeat the subtraction procedure for several generations on the decay chain.
In contrast,
the simultaneous decay curve analysis allows us to determine independent production cross sections of various nuclides by adjustment to the emission rates of all measured $\gamma$ lines at once as demonstrated in the present work.
Unlike determination of a cross section with the activation cross section formula,
we can include in simultaneous decay curve analysis a peak area related with several $\gamma$-lines that are unresolved to each other,
and utilize more information from each spectrum.
It is also very easy to repeat analysis by adding or deleting a product nuclide and/or new decay path in the decay chain model.

We studied the spin cutoff parameter dependence of the $^{198}$Au, $^{197}$Hg and $^{195}$Hg isomeric ratios,
and found that model prediction is improved if the spin cutoff parameter is reduced to the half or less from the one estimated with the rigid body moment of inertia.
This result is consistent with analysis of the mass dependence of $\eta$~\cite{Sudar2018}, our global adjustment of $\eta$ performed to the experimental isomeric ratios~\cite{Rodrigo2023} and another recent global analysis~\cite{Cannarozzo2023}.

We also briefly discussed a question on the quantity specification in the EXFOR library.
The information in an experimental report is often insufficient for EXFOR compilers to distinguish $\sigma_g^\mathrm{cum}$ and $\sigma_g+p\sigma_m$,
and the EXFOR coding rule also does not distinguish them.
The present work is motivated by possible improvement of the current way of compilation,
and we plan to investigate this question further.

The initial $\alpha$-particle beam energy used in the present work is not high enough to study the peak regions of the excitation functions,
and we plan an additional experiment covering higher energies.

\begin{acknowledgement}
This experiment was performed at the RI Beam Factory operated by RIKEN Nishina Center and Center of Nuclear Science, University of Tokyo.
The authors would like to thank Akihiro Nambu, Yudai Shigekawa and Sachiko Usuda for technical assistance with the experiment.
Arjan Koning shared with us TALYS-2.0 for this study prior to its official release.
One of us (NO) thank Valentina Semkova (IAEA) for discussion on interpretation of our decay curves, Tibor Kibedi (Australian National University) for his instruction on usage of the \textsc{gabs} code, and Sophiya Taova (RFNC-VNIIEF) for her communication with Vladimir Utyonkov (JINR) regarding the scale of the excitation functions reported by Sagaidak et al. 
\end{acknowledgement}

\appendix*
\section{Simultaneous decay curve analysis and generalized activation cross section formula}
The formulae required for the simultaneous decay curve analysis and the generalized activation cross section formula are summarized in this appendix.
\subsection{Number of atoms after cooling} 
\label{sec:app1}
If each nuclide on a decay chain 1$\to$2$\to\cdots\to i\to \cdots$ has a single decay branch and there is no production process other than decays,
the number of atoms of each nuclide follows
\begin{equation}
\frac{\del N_i}{\del t}=
\left\{
\begin{array}{ll}
-\lambda_iN_i                         & \mathrm{if\quad} i=1
\\[5mm]
\lambda_{i-1}N_{i-1}-\lambda_iN_i     &\mathrm{otherwise}
\end{array}
\right.,
\label{eqn:Rutherford}
\end{equation}
where $N_i=N_i(t)$ denotes the number of atoms of the nuclide $i$ after cooling time $t$.
Let us assume that only the nuclide \textit{1} presents at $t=0$, namely $N_1(0)=N_1^0$ and $N_i(0)=0$ for $i\ge 2$.
The solution of this differential equation system given by Bateman~\cite{Bateman1910} is
\begin{equation}
N_i(t)=N_1^0 \sum_{k=1}^i \Lambda_k^{1,i} e^{-\lambda_k t}
\label{eqn:formula1}
\end{equation}
with
\begin{equation}
\Lambda_k^{j,i}=
\left\{
\begin{array}{ll}
1 & \mathrm{if\quad} i=j
\\[5mm]
\prod_{l=j}^{i-1}\lambda_l\left/\prod_{\substack{l=j\\(l\ne k)}}^i(\lambda_l-\lambda_k)\right.&\mathrm{otherwise}
\\
\end{array}
\right..
\end{equation}
Examples of the explicit form of $\Lambda_k^{j,i}$ are
\begin{equation}
\begin{array}{lll}
\Lambda_1^{1,2}=\frac{\lambda_1}{\lambda_2-\lambda_1},&
\Lambda_2^{1,2}=\frac{\lambda_1}{\lambda_1-\lambda_2},\\
\Lambda_2^{2,3}=\frac{\lambda_2}{\lambda_3-\lambda_2},&
\Lambda_3^{2,3}=\frac{\lambda_2}{\lambda_2-\lambda_3},\\
\Lambda_1^{1,3}=\frac{\lambda_1\lambda_2}{(\lambda_2-\lambda_1)(\lambda_3-\lambda_1)},&
\Lambda_2^{1,3}=\frac{\lambda_1\lambda_2}{(\lambda_1-\lambda_2)(\lambda_3-\lambda_2)},&
\Lambda_3^{1,3}=\frac{\lambda_1\lambda_2}{(\lambda_1-\lambda_3)(\lambda_2-\lambda_3)}.\\
\end{array}
\end{equation}
\subsection{Number of atoms at end of bombardment}
\label{sec:app2}
Let us extend Eq.~(\ref{eqn:Rutherford}) by adding a production term for the nuclide \textit{1} (production cross section $\sigma_1$) to
\begin{equation}
\frac{\del N_i}{\del t}=
\left\{
\begin{array}{ll}
R_1-\lambda_iN_i                         & \mathrm{if\quad} i=1
\\[5mm]
\lambda_{i-1}N_{i-1}-\lambda_iN_i     &\mathrm{otherwise}
\end{array}
\right.,
\label{eqn:Rutherford2}
\end{equation}
where the reaction rate $R_1=\phi n \sigma_1$ is introduced with the beam flux $\phi$ and sample atom areal number density $n$.
The solution of this differential equation is 
\begin{equation}
N_i(t)=R_1 \sum_{k=1}^i \Lambda_k^{1,i} h_k(t)
\label{eqn:formula3}
\end{equation}
with $h_k(t)=(1-e^{-\lambda_k t})/\lambda_k$.
For example,
\begin{equation}
\begin{array}{lcl}
N_1&=& R_1h_1,
\\[5mm]
N_2&=& R_1 \left[
                  \frac{\lambda_1}{\lambda_2-\lambda_1}h_1
                 +\frac{\lambda_1}{\lambda_1-\lambda_2}h_2
                           \right],
\\[5mm]
N_3&=&R_1 \left[
         \frac{\lambda_1\lambda_2}{(\lambda_2-\lambda_1)(\lambda_3-\lambda_1)} h_1
        +\frac{\lambda_1\lambda_2}{(\lambda_1-\lambda_2)(\lambda_3-\lambda_2)} h_2
        +\frac{\lambda_1\lambda_2}{(\lambda_1-\lambda_3)(\lambda_2-\lambda_3)} h_3
                           \right].
\end{array}
\label{eqn:formula3example}
\end{equation}
\begin{proof}
The number of atoms of the nuclide $i$ produced during short interval $\del \tau$ is $\del \tau R_i$.
According to Eq.~(\ref{eqn:formula1}),
the number of the nuclide produced during $\del \tau$ and survived after cooling for $\tau$ is~\cite{Tasaka1980}
\begin{equation}
\del N_i=\del \tau R_1 \sum_{k=1}^i \Lambda_k^{1,i} e^{-\lambda_k \tau}.
\end{equation}
If the bombardment continues for $t$, 
\begin{equation}
N_i=R_1 \sum_{k=1}^i \Lambda_k^{1,i} \int_0^t \del \tau\, e^{-\lambda_k \tau}
   =R_1 \sum_{k=1}^i \Lambda_k^{1,i} \frac{1-e^{-\lambda_k t}}{\lambda_k}.
\end{equation}
\end{proof}

\subsection{Number of atoms after bombardment and cooling}
\label{sec:app3}
The number of atoms of the nuclide $i$ produced by bombardment for $t_b$ at the reaction rate $R_1=\phi n \sigma_1$ followed by cooling for $t$ is 
\begin{equation}
N_i(t)=R_1 \sum_{k=1}^i \Lambda_k^{1,i} g_k(t)
\label{eqn:formula4}
\end{equation}
with $g_k(t)=(1-e^{-\lambda_k t_b})e^{-\lambda_k t}/\lambda_k$.
For example,
\begin{equation}
\begin{array}{lcl}
N_1&=&R_1\,g_1,
\\[5mm]
N_2&=&R_1\left[\frac{\lambda_1}{\lambda_2-\lambda_1}g_1+\frac{\lambda_1}{\lambda_1-\lambda_2}g_2\right],
\\[5mm]
N_3&=&R_1\left[
      \frac{\lambda_1\lambda_2}{(\lambda_2-\lambda_1)(\lambda_3-\lambda_1)}g_1
     +\frac{\lambda_1\lambda_2}{(\lambda_1-\lambda_2)(\lambda_3-\lambda_2)}g_2
     +\frac{\lambda_1\lambda_2}{(\lambda_1-\lambda_3)(\lambda_2-\lambda_3)}g_3
        \right].
\\
\end{array}
\label{eqn:formula4example}
\end{equation}

\begin{proof}
Considering the relation $g_k(0)=h_k(t_b)$,
it is obvious that Eq.~(\ref{eqn:formula4}) satisfies the initial condition given by Eq.~(\ref{eqn:formula3}) at $t=0$.
Furthermore,
we can demonstrate Eq.~(\ref{eqn:formula4}) and 
\begin{equation}
N_{i+1}(t)=R_1 \sum_{k=1}^{i+1} \Lambda_k^{1,i+1} g_k(t)
\label{eqn:formula4prediction1}
\end{equation}
satisfy
\begin{equation}
\frac{\del N_{i+1}}{\del t}=-\lambda_{i+1}N_{i+1}+\lambda_iN_i
\label{eqn:formula4equation}
\end{equation}
as follows:
\begin{eqnarray}
& &
\frac{\del N_{i+1}}{\del t}+\lambda_{i+1}N_{i+1}-\lambda_iN_i
\nonumber\\
&=&
              R_1\sum_{k=1}^{i+1}\Lambda_k^{1,i+1}\frac{1-e^{-\lambda_k t_b}}{\lambda_k}(-\lambda_k)e^{-\lambda_k t}
+\lambda_{i+1}R_1\sum_{k=1}^{i+1}\Lambda_k^{1,i+1}\frac{1-e^{-\lambda_k t_b}}{\lambda_k}e^{-\lambda_k t}
 -\lambda_i   R_1\sum_{k=1}^i    \Lambda_k^{1,i}  \frac{1-e^{-\lambda_k t_b}}{\lambda_k}e^{-\lambda_k t}
\nonumber\\
&=&           R_1\sum_{k=1}^{i+1}\Lambda_k^{1,i+1}\left(\frac{\lambda_{i+1}}{\lambda_k}-1\right)(1-e^{-\lambda_k t_b})e^{-\lambda_k t}
             -R_1\sum_{k=1}^i\Lambda_k^{1,i}\frac{\lambda_i}{\lambda_k}(1-e^{-\lambda_k t_b})e^{-\lambda_k t}
\nonumber\\
&=&           R_1\sum_{k=1}^i \Lambda_k^{1,i+1}\left(\frac{\lambda_{i+1}}{\lambda_k}-1\right)(1-e^{-\lambda_k t_b})e^{-\lambda_k t}
             -R_1\sum_{k=1}^i\Lambda_k^{1,i}\frac{\lambda_i}{\lambda_k}(1-e^{-\lambda_k t_b})e^{-\lambda_k t}
=0,
\end{eqnarray}
where the relation $\Lambda_k^{1,i+1}=\Lambda_k^{1,i}\lambda_i/(\lambda_{i+1}-\lambda_k)$ $(k\le i)$ is used.
\end{proof}
\subsection{Number of decays during measurement after bombardment and cooling}
\label{sec:app4}
The number of decays of the nuclide $i$ during time interval $t$ after bombardment for $t_b$ at the reaction rate $R_1$ and after cooling for $t_c$ is 
\begin{equation}
C_i(t)=\lambda_i R_1 \sum_{k=1}^i \Lambda_k^{1,i} \frac{f_k(t)}{\lambda_k},
\label{eqn:formula5}
\end{equation}
where
$f_k(t)=(1-e^{-\lambda_k t_b})e^{-\lambda_k t_c}(1-e^{-\lambda_k t})/\lambda_k$.
For example,
\begin{equation}
\begin{array}{lcl}
C_1&=&R_1\,f_1,
\\[5mm]
C_2&=&R_1\left[
      \frac{\lambda_2}{\lambda_2-\lambda_1}f_1
     +\frac{\lambda_1}{\lambda_1-\lambda_2}f_2
         \right],
\\[5mm]
C_3&=&R_1\left[
     \frac{\lambda_2\lambda_3}{(\lambda_2-\lambda_1)(\lambda_3-\lambda_1)}f_1
   +\frac{\lambda_1\lambda_3}{(\lambda_1-\lambda_2)(\lambda_3-\lambda_2)}f_2
   +\frac{\lambda_1\lambda_2}{(\lambda_1-\lambda_3)(\lambda_2-\lambda_3)}f_3
\right].
\end{array}
\label{eqn:formula5example}
\end{equation}
\begin{proof}
According to Eq.~(\ref{eqn:formula4}),
the decay rate of the nuclide $i$ after bombardment for $t_b$ and cooling for $t_c+\tau$ is
\begin{equation}
-\frac{\del N_i}{\del t}=\lambda_i N_i = \lambda_i R_1 \sum_{k=1}^i \Lambda_k^{1,i} \frac{(1-e^{-\lambda_k t_b})e^{-\lambda_k (t_c+\tau)}}{\lambda_k}.
\end{equation}
The number of atoms of the nuclide $i$ decaying during the short interval $\del \tau$ after measurement for $t_c+\tau$ is
\begin{equation}
\del C_i=\del \tau \lambda_i N_i =\del \tau \lambda_i R_1 \sum_{k=1}^i \Lambda_k^{1,i} \frac{(1-e^{-\lambda_k t_b})e^{-\lambda_k (t_c+\tau)}}{\lambda_k}.
\end{equation}
The number of the nuclide decays between $t_c$ and $t_c+t$ is
\begin{eqnarray}
C_i&=&\lambda_i R_1 \sum_{k=1}^i \Lambda_k^{1,i} \frac{1-e^{-\lambda_k t_b}}{\lambda_k}
   e^{-\lambda_k t_c}
   \int_0^{t} \del \tau\,e^{-\lambda_k \tau}
   =\lambda_i R_1 \sum_{k=1}^i \Lambda_k^{1,i} \frac{(1-e^{-\lambda_k t_b})e^{-\lambda_k t_c}(1-e^{-\lambda_k t})}{\lambda_k^2}
.
\end{eqnarray}
\end{proof}

\subsection{Generalization}
\label{sec:app5}
The equations developed so far can be easily generalized to a case where not only the nuclide \textit{1} but also nuclides $j$ $(j=2,i)$ are produced with reaction rate $R_j$.
Assuming that the nuclide $j$ has only one mother nuclide $j-1$,
the number of atoms of the nuclide $i$ after bombardment and cooling is generalized from Eq.~(\ref{eqn:formula4}) to
\begin{equation}
N_i(t)=\sum_{j=1}^i R_j p_{ji} \sum_{k=j}^i \Lambda_k^{j,i} g_k(t),
\label{eqn:formula4gen}
\end{equation}
and the number of decays of the nuclide $i$ during measurement for $t$ after bombardment for $t_b$ and cooling for $t_c$ is generalized from Eq.~(\ref{eqn:formula5}) to
\begin{equation}
C_i(t)=\lambda_i \sum_{j=1}^i R_j p_{ji} \sum_{k=j}^i \Lambda_k^{j,i} \frac{f_k(t)}{\lambda_k},
\label{eqn:formula5gen}
\end{equation}
where $p_{ji}$ is the branching ratio for decay from the nuclide $j$ to nuclide $i$ and $R_j=\phi n \sigma_j$.
These equations can be further generalized to a case where several decay paths from $j$ to $i$ exists (c.f. two decay paths from $^{195m}$Pt to $^{195}$Au discussed in the next subsection).

\subsection{Decay curve analysis}
\label{sec:app6}
\begin{figure}
\begin{center}
\centerline{\includegraphics[scale=0.4]{"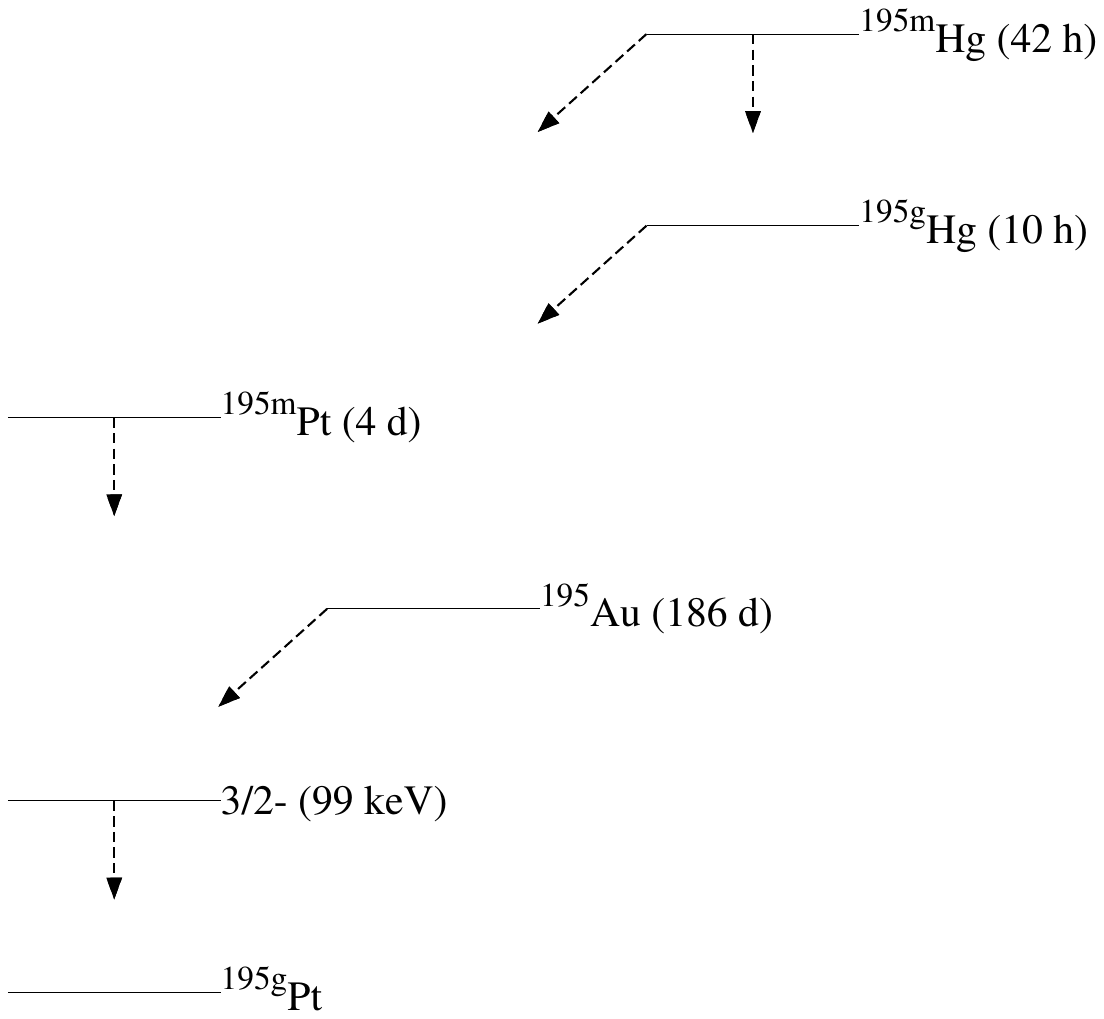"}}
\end{center}
\caption{$A=195$ decay chain relevant to 99~keV $\gamma$-ray emission from a platinum foil irradiated by an $\alpha$ particle beam at 29~MeV.}
\label{fig:decayscheme99keV}
\end{figure}

Here we demonstrate application of Eq.~(\ref{eqn:formula4example}) to the decay curve analysis of 99~keV $\gamma$-rays of a platinum foil irradiated by an $\alpha$-particle beam for 1 h at 29~MeV.
This $\gamma$ line corresponds to the transition from the first excitation state to the ground state of $^{195}$Pt following the electron capture (EC) of $^{195}$Au (186~d) or isometric transition (IT) of $^{195m}$Pt (4~d).
The decay scheme of the nuclides contributing to the 99~keV $\gamma$-ray emission is illustrated in Fig.~\ref{fig:decayscheme99keV},
which shows that emission of the 99~keV $\gamma$-rays originates from the production of $^{195}$Au and $^{195m}$Pt as well as production of $^{195g}$Hg (10~h) and $^{195m}$Hg (42~h).
For simplicity, we assume that production of $A=195$ nuclides with $Z\le 77$ is negligible.
If we denote $N_1$ to $N_3$ in Eq.~(\ref{eqn:formula4example}) by
$N_1=N_1(t,R_1,\lambda_1)$,
$N_2=N_2(t,R_1,\lambda_1,\lambda_2)$ and
$N_3=N_3(t,R_1,\lambda_1,\lambda_2,\lambda_3)$,
the partial emission rates of the 99~keV $\gamma$-rays originating from the production and decay routes are 
\begin{equation}
\begin{array}{ll}
I_A\,\lambda_A\,p_{\mathrm{IT}}\,N_3(t,R_m,\lambda_m,\lambda_g,\lambda_A) &\text{for $^{195m}$Hg $\to$ $^{195g}$Hg $\to$ $^{195}$Au $\to$ $^{195g}$Pt},\\
I_A\,\lambda_A\,p_{\mathrm{EC}}\,N_2(t,R_m,\lambda_m,\lambda_A)           &\text{for $^{195m}$Hg $\to$ $^{195}$Au $\to$ $^{195g}$Pt                  },\\
I_A\,\lambda_A\,                 N_2(t,R_g,\lambda_g,\lambda_A)           &\text{for $^{195g}$Hg $\to$ $^{195}$Au $\to$ $^{195g}$Pt                  },\\
I_A\,\lambda_A\,                 N_1(t,R_A,\lambda_A)                     &\text{for $^{195}$Au  $\to$ $^{195g}$Pt                                   },\\
I_P\,\lambda_P\,                 N_1(t,R_P,\lambda_P)                     &\text{for $^{195m}$Pt $\to$ $^{195g}$Pt                                   },\\
\end{array}
\label{eqn:partialrate}
\end{equation}
where 
the suffices $A$, $P$, $m$ and $g$ stand for $^{195}$Au, $^{195m}$Pt, $^{195m}$Hg and $^{195g}$Hg, respectively,
$I$ is the $\gamma$ emission probability, 
and
$p_\mathrm{IT}$ and $p_\mathrm{EC}$ stand for the IT and EC branching ratios of $^{195m}$Hg.

The left-top panel of Fig.~\ref{fig:decaycurve} shows cooling time dependence of the emission rate for the 99~keV $\gamma$-rays.
By fitting the sum of the five partial emission rates expressed by Eq.~(\ref{eqn:partialrate}) to the measured total emission rate,
one can determine the reaction rates for production of $^{195m}$Hg, $^{195g}$Hg, $^{195}$Au and $^{195m}$Pt.
Their production cross sections can be easily obtained by dividing the reaction rates by the beam flux and sample atom areal number density.

\subsection{Generalized activation cross section formula}
\label{sec:app7}
If the nuclides on the decay chain 1$\to$2$\to\cdots\to i$ can be produced with production cross sections $\sigma_j$ ($j=1,i$),
and each nuclide has only one decay precursor,
$\sigma_i$ is related with the number of decays $C_j$ ($j=1,i$) by 
\begin{equation}
\sigma_i=
\left\{
\begin{array}{ll}
\sigma_i^\mathrm{cum}
&\mathrm{if\quad} i=1
\\
\sigma_i^\mathrm{cum}
   -\frac{\lambda_i}{f_i}\sum_{j=1}^{i-1}\sigma_jp_{ji}\sum_{k=j}^i \Lambda_k^{j,i}\frac{f_k}{\lambda_k}
&\mathrm{otherwise}
\end{array}
\right.
\label{eqn:formula6}
\end{equation}
with
$\sigma_i^\mathrm{cum}=C_i/(f_i \phi n)$ and the decay branching ratio from the nuclide $j$ to nuclide $i$ denoted by $p_{ji}$.
For example,
\begin{equation}
\begin{array}{lcl}
\sigma_1&=&\sigma_1^\mathrm{cum},
\\[5mm]
\sigma_2&=&\sigma_2^\mathrm{cum}- \sigma_1p_{12}\left[\frac{f_1}{f_2}\frac{\lambda_2}{\lambda_2-\lambda_1}+\frac{\lambda_1}{\lambda_1-\lambda_2}\right],
\\[5mm]
\sigma_3&=&\sigma_3^\mathrm{cum}- \sigma_2p_{23}\left[\frac{f_2}{f_3}\frac{\lambda_3}{\lambda_3-\lambda_2}+\frac{\lambda_2}{\lambda_2-\lambda_3}\right]
\\[5mm]
   &-&  \sigma_1p_{13}\left[\frac{f_1}{f_3}\frac{\lambda_2\lambda_3}{(\lambda_2-\lambda_1)(\lambda_3-\lambda_1)}\right.
    +       \frac{f_2}{f_3}\frac{\lambda_1\lambda_3}{(\lambda_1-\lambda_2)(\lambda_3-\lambda_2)}
    + \left.\frac{\lambda_1\lambda_2}{(\lambda_1-\lambda_3)(\lambda_2-\lambda_3)}\right].
\end{array}
\end{equation}
\begin{proof}
The formula for $i=1$ is obvious from Eq.~(\ref{eqn:formula5example}).
For $i\ge2$, 
Eq.~(\ref{eqn:formula5gen}) gives
\begin{equation}
C_i=\lambda_i \sum_{j=1}^i R_j p_{ji} \sum_{k=j}^i \Lambda_k^{j,i} \frac{f_k}{\lambda_k} = R_i f_i + \lambda_i \sum_{j=1}^{i-1} R_j p_{ji} \sum_{k=j}^i \Lambda_k^{j,i} \frac{f_k}{\lambda_k},
\end{equation}
where $\Lambda_i^{i,i}=1$ is used.
This is solved in terms of $R_i$ to
\begin{equation}
R_i=\frac{C_i}{f_i}
   -\frac{\lambda_i}{f_i}\sum_{j=1}^{i-1} R_j p_{ji}\sum_{k=j}^i \Lambda_k^{j,i}\frac{f_k}{\lambda_k}
   =R_i^\mathrm{cum}
   -\frac{\lambda_i}{f_i}\sum_{j=1}^{i-1} R_j p_{ji}\sum_{k=j}^i \Lambda_k^{j,i}\frac{f_k}{\lambda_k},
\end{equation}
where $R_i^\mathrm{cum}=C_i/f_i$.
By dividing the both sides of this equation by $\phi n$,
we obtain Eq.~(\ref{eqn:formula6}) with $i\ge 2$.
\end{proof}

\bibliography{Pt+a}
\end{document}